\documentclass[5p,times,authoryear]{elsarticle}

\usepackage{lineno}
\modulolinenumbers[5]
\journal{Electronic Commerce Research and Applications}

\usepackage{graphicx,color,colortbl,psfrag,subfigure}
\usepackage[usenames,dvipsnames]{xcolor}
\usepackage{bbding}
\usepackage{subfigure}
\usepackage{tabularx}
\usepackage{multirow}
\usepackage{array}
\usepackage[figuresright]{rotating}

\usepackage{amssymb}
\usepackage{amsmath}
\usepackage{amsfonts}

\usepackage[bookmarks=true,
			colorlinks=true,
			linkcolor=black,
			urlcolor=black, 
			citecolor=black]{hyperref}
\usepackage{url}

\usepackage{natbib}
\biboptions{longnamesfirst,semicolon}

\usepackage{bm}
\usepackage{algpseudocode}
\usepackage{algorithm}

\usepackage{pifont}

\begin{document}

\begin{frontmatter}

\title{A Lattice Framework for Pricing Display Advertisement Options with the Stochastic Volatility Underlying Model}

\author[rvt]{Bowei Chen\corref{cor1}}
\ead{bowei.chen@cs.ucl.ac.uk}
\cortext[cor1]{Corresponding author}

\author[rvt]{Jun Wang}
\ead{jun.wang@cs.ucl.ac.uk}

\address[rvt]{University College London, Gower Street, London, WC1E 6DY, United Kingdom}

\begin{abstract}
Advertisement (abbreviated ad) options are a recent development in online advertising. Simply, an ad option is a first look contract in which a publisher or search engine grants an advertiser a right but not obligation to enter into transactions to purchase impressions or clicks from a specific ad slot at a pre-specified price on a specific delivery date. Such a structure provides advertisers with more flexibility of their guaranteed deliveries. The valuation of ad options is an important topic and previous studies on ad options pricing have been mostly restricted to the situations where the underlying prices follow a geometric Brownian motion (GBM). This assumption is reasonable for sponsored search; however, some studies have also indicated that it is not valid for display advertising. In this paper, we address this issue by employing a stochastic volatility (SV) model and discuss a lattice framework to approximate the proposed SV model in option pricing. Our developments are validated by experiments with real advertising data: (i) we find that the SV model has a better fitness over the GBM model; (ii) we validate the proposed lattice model via two sequential Monte Carlo simulation methods; (iii) we demonstrate that advertisers are able to flexibly manage their guaranteed deliveries by using the proposed options, and publishers can have an increased revenue when some of their inventories are sold via ad options.
\end{abstract}


\end{frontmatter}

\section{Introduction}
\label{ao:introduction}

Options have been widely used in many fields: financial options are an important derivative when speculating profits as well as hedging risk~\citep{Wilmott_2006_1}; real options are an effective decision-making tool for business projects valuation and corporate risk management~\citep{Boer_2002}. Recently, options have been introduced into the field of online advertising to solve the so called non-guaranteed delivery problem as well as to provide advertisers with greater flexibility in purchasing premium ad inventories. \cite{Moon_2010} proposed an ad option for advertisers to make a flexible choice of payment at either cost-per-mille (CPM) or cost-per-click (CPC). They are two popular online advertising payment schemes: the former allows an advertiser to pay when his ad is displayed 1000 times to online users while with the latter an advertiser pays only when his ad is clicked by an online user. The proposal of~\cite{Moon_2010} was similar to an \emph{option paying the worst and cash}~\citep{Zhang_1998} because the option payoff depends on the minimum difference between CPM and CPC. \citet*{Wang_2012_1} proposed a simple European ad option between buying and non-buying the impressions that will be created in the future, and discussed the option pricing based on the \emph{one-step binomial lattice method}~\citep*{Sharpe_1978}. Their ad option was priced from the perspective of a risk-averse publisher who wants to hedge the expected revenue in the next step. \citet*{Chen_2015_1} investigated a special option for sponsored search whereby an advertiser can target a set of keywords for a certain number of total clicks in the future. Each candidate keyword can be specified with a fixed payment price and the option buyer can exercise the option multiple times at any time prior to or on the contract expiration date. Their design was a generalization of the \emph{dual-strike call option}~\citep{Zhang_1998} and the \emph{multi-exercise option}~\citep{Marshall_2012}.

\begin{table*}[t]
\centering
\caption{Summary of lattice methods used in pricing a call option written on an asset with the GBM underlying model. Detailed description of notations is provided in Table~\ref{tab:notation_summary}.}
\label{tab:lattice_method_review}
\begin{tabular}{p{1in}|p{2.7in}|p{2.5in}}
\hline
Model 
& 
Movement scales $u, d$ (or $u, m, d$) 
& 
Transition probabilities $q_1, q_2, \cdots q_k$\\
\hline
\multicolumn{3}{l}{
Binomial lattice (one factor)
}
\\
\hline
CRR 
&
$u = e^{\sigma \sqrt{\Delta t}}$, $d = 1/u$. 
& 
$q_1 = \frac{e^{r \Delta t} - d}{u - d}$, 
$q_2 = 1-q_1$.
\\
\hline
Tian-BIN
& 
$u = \frac{\gamma\zeta}{2}(\zeta + 1 + \sqrt{\zeta^2 + 2\zeta - 3})$,
$\gamma = e^{r \Delta t}$,
& 
$q_1 = \frac{e^{r \Delta t} - d}{u - d}$, 
$q_2 = 1-q_1$.
\\
&
$d = \frac{\gamma\zeta}{2}(\zeta + 1 - \sqrt{\zeta^2 + 2\zeta - 3})$,
$\zeta = e^{\sigma^2 \Delta t}$.
&
\\
\hline
Haahtela-BIN
&
$u = e^{\sqrt{e^{\sigma^2 \Delta t} - 1} + r \Delta t}$, 
$d = e^{- \sqrt{e^{\sigma^2 \Delta t} - 1} + r \Delta t}$.  
&
$q_1 = \frac{e^{r \Delta t} - d}{u - d}$, 
$q_2 = 1-q_1$.
\\
\hline
\multicolumn{3}{l}{
Trinomial lattice (one factor)
}
\\
\hline
Boyle-TRIN
&
$u = e^{\lambda \sigma \sqrt{\Delta t}}$,
&
$q_1 = \frac{(\zeta + \gamma^2 - \gamma) u - (\gamma - 1)}{(u - 1)(u^2 - 1)}$,
\\
&
$m = 1$, 
&
$q_2 = 1 - q_1 - q_3$, $\zeta = e^{2 r \Delta t} \big( e^{\sigma^2 \Delta t} - 1\big)$,
\\
&
$d = e^{- \lambda \sigma \sqrt{\Delta t}}$.
&
$q_3 = \frac{(\zeta + \gamma^2 - \gamma) u^2 - (\gamma - 1) u^3}{(u - 1)(u^2 - 1)}$, $\gamma = e^{r \Delta t}$.
\\
\hline
KR-TRIN
&
$u = e^{\lambda \sigma \sqrt{\Delta t}}$,
&
$q_1 = \frac{1}{2 \lambda^2} + \frac{(r - \frac{1}{2}\sigma^2)\sqrt{\Delta t}}{2 \lambda \sigma}$,
\\
&
$m = 1$, 
&
$q_2 = 1 - \frac{1}{\lambda^2}$,
\\
&
$d = e^{- \lambda \sigma \sqrt{\Delta t}}$.
&
$q_3 = \frac{1}{2 \lambda^2} - \frac{(r - \frac{1}{2}\sigma^2) \sqrt{\Delta t}}{2 \lambda \sigma}$.
\\
\hline
Tian-TRIN 
&
$u = \varpi + \sqrt{\varpi^2 - m^2}$,
&
$q_1 = \frac{md - \gamma (m + d) + \gamma^2 \zeta}{(u -d)(u - m)}$,
\\
&
$m = \gamma \zeta^2$, $\gamma = e^{r \Delta t}$, 
$\zeta = e^{\sigma^2 \Delta t}$,
&
$q_2 = \frac{\gamma (u + d) - ud - \gamma^2 \zeta}{(u - m)(m - d)}$,
\\
&
$d = \varpi - \sqrt{\varpi^2 - m^2}$, $\varpi = \frac{\gamma}{2} (\zeta^4 + \zeta^3)$.
&
$q_3 = \frac{um - \gamma (u + m) + \gamma^2 \zeta}{(u - d)(m - d)}$.
\\
\hline
\multicolumn{3}{p{6.75in}}{
Note: CRR~\citep{Cox_1979}; Tian-BIN and Tian-TRIN~\citep{Tian_1993}; Haahtela-BIN~\citep{Haahtela_2010}; Boyle-TRIN~\citep{Boyle_1988}; and KR-TRIN~\citep{Kamrad_1991}.
}
\\
\hline
\end{tabular}
\end{table*}

In this paper, we discuss an ad option that gives an advertiser a right but not obligation to purchase the future impressions or clicks from a specific ad slot (or user tag or keyword) at a pre-specified price. The pre-specified price is also called the \emph{strike price}, which can be same or different to the payment scheme of its underlying ad format. For example, the underlying price (i.e., the winning payment price) of a display impression from real-time bidding (RTB) is usually measured by CPM while the proposed ad option can be specified with a strike price in terms of CPC for this impression. The publisher or search engine who grants this right in exchange for a certain amount of upfront fee, is called the \emph{option price}. Obviously, ad options are more flexible than guaranteed contracts~\citep{Bharadwaj_2010} as on the delivery date. If the advertiser thinks that the spot market is more beneficial, he can join RTB as a bidder and his cost of not using an ad option is only the option price. A contract with a such structure is also called a \emph{first look at inventory} (shortly first look) contract or tactic~\citep{IAB_2015}. It means that an advertiser is given the opportunity to buy inventories which a publisher offers to him, and if he has no use for it, it can be sold onto another ad network. The ad options proposed by our this study and~\cite{Wang_2012_1}, and \cite{Chen_2015_1} are first look contracts while the ad option studied by~\cite{Moon_2010} is not a first look contract.

When pricing an ad option, the previous research is mostly restricted in their usage to those situations where the underlying price follows a \emph{geometric Brownian motion (GBM)}~\citep{Samuelson_1965_2}. According to~\cite{Yuan_2013_2}, \cite{Yuan_2014} and \cite{Chen_2014_2}, there is only a very small number of ad inventories whose CPM or CPC satisfies this assumption. Therefore, the previous studies fail to provide an effective unified framework that covers general situations. In this paper, we address the issue and provide a more general pricing framework. We use a \emph{stochastic volatility (SV) model} to describe the underlying price movement for cases where the GBM assumption is not valid. Based on the SV model, a censored binomial lattice is then constructed for option pricing. We also examine several previous binomial and trinomial lattice methods to price an ad option whose underlying inventory prices follow a GBM model, and deduce the close-form solutions to examine their convergence performance. Our developments are validated by experiments using real advertising data. We examine the fitness of the underlying model, valid the proposed option pricing method, and illustrate that the options provide a more flexible way of selling and buying ads. In particular, we show that an advertiser can have better deliveries in a bull market (where the underlying price increases). On the other hand, a publisher or search engine is able to reduce the revenue volatility over time. In a bear market (where the underlying price decreases), there is a growth in total revenue. To our best knowledge, this is the first work that discusses lattice methods for the ad option evaluation.

The rest of the paper is organized as follows. Section~\ref{ao:related_work} reviews the related work. Section~\ref{ao:preliminaries_lattice} introduces the preliminaries of lattice methods for pricing an ad options with the GBM underlying model. Section~\ref{ao:censored_binomial_lattice} discusses our lattice method to price an ad option with the SV underlying model. Section~\ref{ao:experiments} presents our experimental results. Section~\ref{ao:conclusion} concludes the paper.

\section{Literature Review}
\label{ao:related_work}

The ad options discussed in this paper are closely connected to financial options, whose evaluation can be traced back to~\citet*{Bachelier_1900}, who proposed to use a continuous-time random walk as the underlying process to price an option written on a stock. \citet*{Samuelson_1965_2} then replaced the Bachelier's assumption with a geometric form, called the \emph{geometric Brownian motion (GBM)}. Based on the GBM, \citet*{Black_1973} and~\citet*{Merton_1973} discussed a risk-neutral option pricing method independently, called the \emph{Black-Scholes-Merton (BSM) model}, opening the floodgates to option pricing. Various numerical procedures have appeared in this field, including lattice methods, finite difference methods, Monte Carlo simulations, etc. These numerical procedures are capable of evaluating more complex options when the close-form solution does not exist. In our discussion, we focus on lattice methods. 

\citet*{Sharpe_1978} initiated the concept of pricing a call option written on an asset with simple up and down two-state price changes. We call this the \emph{one-step binomial lattice method} and use it as a pedagogical framework to explain the continuous-time option pricing model without reference to stochastic calculus. \citet*{Cox_1979}~then developed a multi-step binomial framework, called the \emph{Cox-Ross-Rubinstein (CRR) model}, which can converge with the BSM model if the length of the time step is sufficiently small. \citet*{Boyle_1986}~proposed a trinomial lattice, whereby the asset price can either move upwards, downwards, or stay unchanged in a given time period. Other contributors to one factor lattice methods include~\citet*{Kamrad_1991}, \citet*{Tian_1993} and~\citet*{Haahtela_2010}. The technical details and differences of these methods are presented in Table~\ref{tab:lattice_method_review}, where the \emph{movement scale} is the ratio of the price in the next state to the current one, and the \emph{transition probability} is the risk-neutral probability that the asset price moves from the current state to the next one, which is labelled from the upper state to the lower state. It is also worth noting that all of these methods adopt Samuelson's GBM assumption for the underlying asset price.  

The GBM assumption may not always be valid empirically. This motivates a general Ornstein-Uhlenbeck (OU) diffusion process for option pricing. \citet*{Nelson_1990} discussed the conditions under which a sequence of binomial processes converges weakly to an OU diffusion process and investigated its application to pricing an option written on an asset with constant volatility. \citet*{Primbsa_2007} then proposed a pentanomial lattice method that incorporates the skewness and kurtosis of the underlying asset price and found that the limiting distribution is compounded Poisson. \citet*{Nelson_1990} and \citet*{Primbsa_2007} solved the lattice pricing for the non-GBM underlyings which have constant volatility. \citet{Florescu_2005,Florescu_2008}~proposed lattice methods that deal with a general SV underlying model. However, their method is not very practical in terms of computational efficiency as the transition probabilities are restricted by many conditions and need to be estimated independently before building up the price lattice. From our point of view, a direct censor on transition probabilities of each node, as discussed in~\citep*{Nelson_1990}, would be more efficient. Our proposed method in Section~\ref{ao:censored_binomial_lattice} is based on this idea.

\section{Preliminaries of Lattice Method}
\label{ao:preliminaries_lattice}

\begin{table}[t]
\centering
\caption{Summary of key notations and abbreviations.}
\label{tab:notation_summary}
\begin{tabular}{p{0.5in}|p{2.6in}}
\hline
Notation & Description \\
\hline
$T$             & Option expiration date (in terms of year).\\
$n$             & Total number of time steps and the length of each time step is $\Delta t = T/n$.\\
$\widehat{r}, \widetilde{r}, r$   & Constant risk-less interest rate: $\widehat{r}$ is the discrete-time interest rate in $\Delta t$; $\widetilde{r} = 1+ \widehat{r}$; and $r$ is a continuous-time interest rate where $e^{r \Delta t} = \widetilde{r}$.
\\
$u, m, d$   & State transition size (or movement scale) in upward, unchanged and downward movement. \\ 
$q_1,\ldots,q_n$ & Risk-neutral state transition probability, labelled from the top node to the bottom node.\\
$Q^{\{i\}}(t_k)$ & Risk-neutral probability on node $i$ at time $t_k$.\\
$\mathbb{Q}$ & Risk-neutral probability measure. \\
$\mathbb{P}$ & Real-world probability measure. \\
$M_i$       & $M_i$ is CPM at time step $i$, $i = 0, \ldots, n$.\\
$ M(t)$     & $M(t)$ is  CPM at time $t$. \\
$C_i$       & $C_i$ is CPC at time step $i$.\\ 
$C(t)$      & $C(t)$ is CPC at time $t$.\\
$H$       & Constant CTR.\\
$\Phi_n$    & Option payoff on the expiration date.\\
$F^M, F^C$  & Strike price in terms of CPM, CPC.\\
$\pi_0$     & Option price at time $0$ (i.e., the time step $0$).\\
$\mathcal{N}(\cdot)$ & Cumulative distribution function of a standard normal distribution.\\
$\mathbf{N}(x, y^2)$ & Normal distribution with mean $x$ and standard deviation $y$, where $x, y \in \mathbb{R}$. \\ 
$\mu$ & Constant drift for the underlying price.\\
$\sigma$ & Constant volatility of the underlying price. \\
$\sigma(t)$  & Stochastic volatility of the underlying price.\\
$\kappa, \theta, \delta$ & Constant speed, long-term mean, and volatility for the stochastic volatility model.\\
CPC             & Cost-per-click.\\
CPM             & Cost-per-mille (i.e., 1000 impressions).\\
CTR			    & Click-through rate.\\
$\mathbb{E}[\cdot]$ & Expectation.\\
$\mathrm{std}(\cdot)$ & Standard deviation.\\
$x \wedge y$ & $\min\{x, y\}$, where $x, y \in \mathbb{R}$.\\
$(\cdot)^+$  & $\max\{0, \cdot\}$.\\
\hline
\end{tabular}
\end{table}

This section introduces the basic settings of the lattice based option pricing framework in the context of online advertising. The previous lattice methods introduced in Table~\ref{tab:lattice_method_review} are examined. For the reader's convenience, the key notations and terminologies used throughout the paper are described in Table~\ref{tab:notation_summary}. We here discuss the case where an ad option allows its buyer to pay a fixed CPC for display impressions. Therefore, the strike price of the option is the fixed CPC and the underlying price is the uncertain winning payment CPM from RTB, where each single impression being auctioned off is paid at the second highest bid~\citep{Edelman_2007_2,Google_2011}. Other ad option cases can be discussed in the same manner, such as an ad option allows its buyer to pay a fixed CPM for display impressions, or an ad option allows its buyer to pay a fixed CPM or CPC for clicks.

Suppose that an advertiser buys an ad option in time $0$ which allows him to purchase several impressions from a publisher's ad slot in time $1$ at a fixed CPC, denoted by $F^C$. As impressions are normally auctioned off at a CPM value, the underlying price is the winning payment CPM from RTB, denoted by $M_i$, $i=0,1$. In time $1$, the underlying price may rise or fall, denoted by $M_1^{\{u\}}$ or $M_1^{\{d\}}$. Let us consider the upward case. If $M_1^{\{u\}}/(1000 H) \geq F^C$, the advertiser will exercise the option; if $M_1^{\{u\}}/(1000 H) < F^C$, he will not exercise the option but join RTB instead. Note that $H$ represents a constant CTR; therefore, the underlying and strike prices can be compared on the same measurement basis. Mathematically, we use the option payoff function $\Phi_1^{\{u\}}$ to describe the above decision making, $\Phi_1^{\{u\}} := (M_1^{\{u\}}/(1000 H)  - F^C)^+$. Similarly, if the winning payment CPM moves downward, the option payoff $\Phi_1^{\{d\}} := (M_1^{\{d\}}/(1000 H)  - F^C)^+$.

\begin{figure*}[t]
\centering
\includegraphics[width=0.8\linewidth]{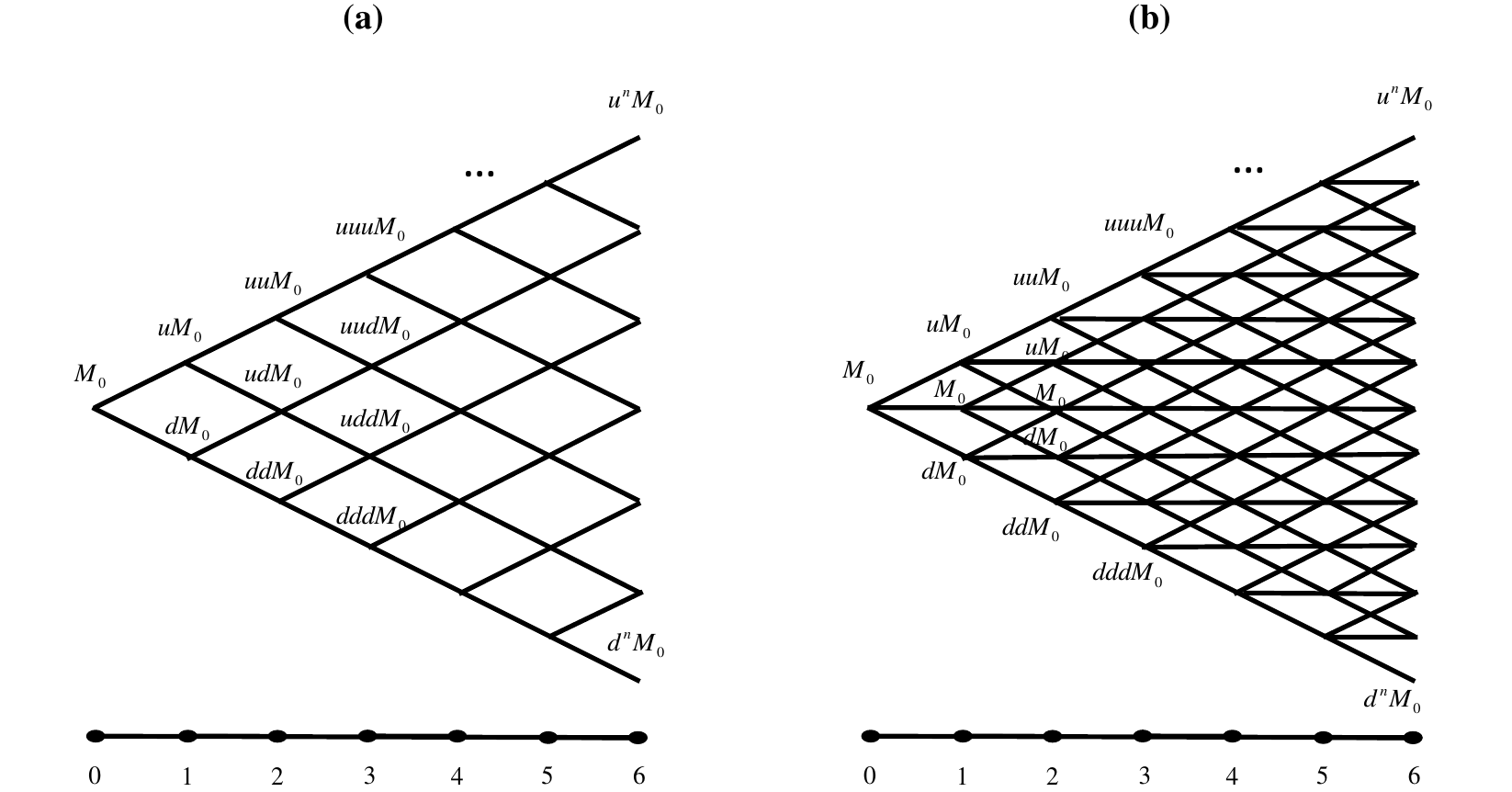}
\caption{Lattice framework: (a) the binomial lattice for CPM; (b) the trinomial lattice for CPM.}
\label{fig:lattice_framework}
\includegraphics[width=0.8\linewidth]{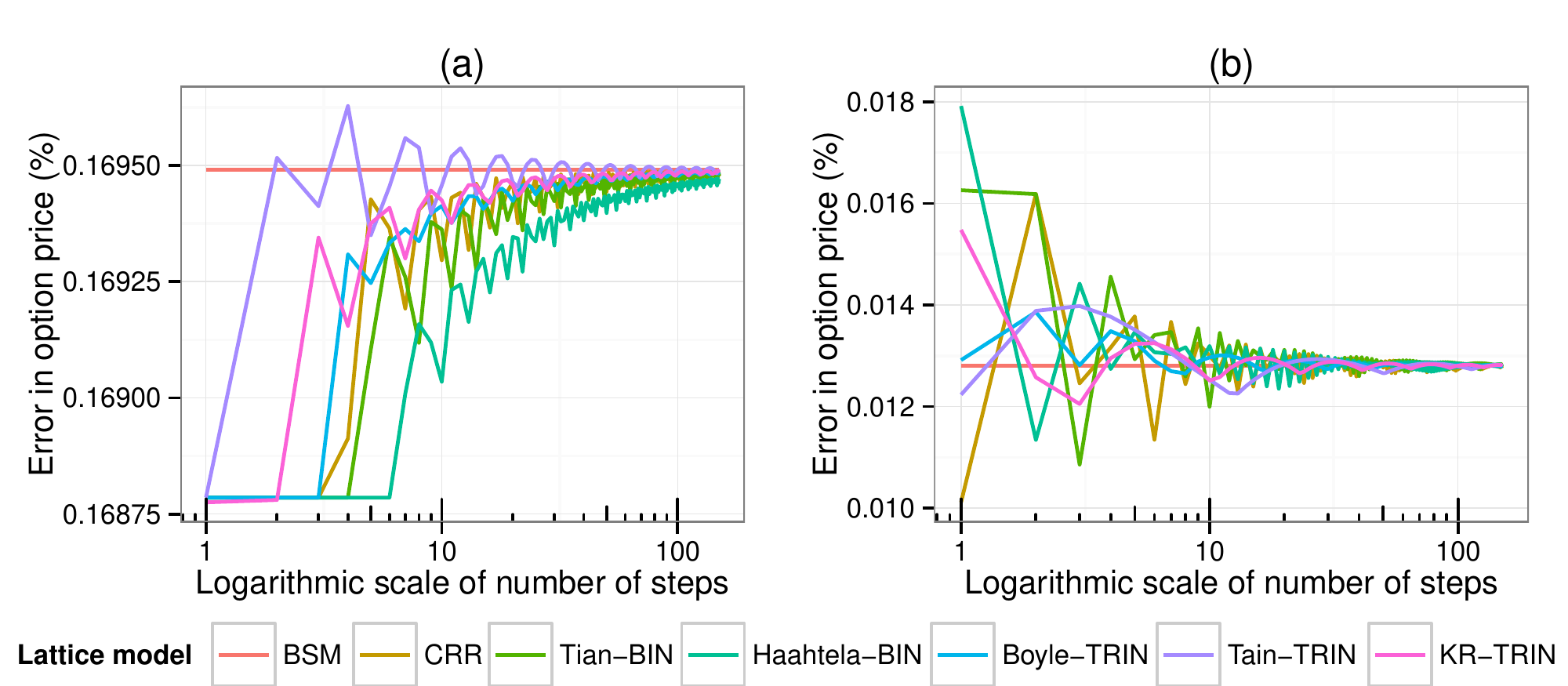}
\caption{Comparison of the convergence performance of the binomial and trinomial lattice methods for pricing a display ad option with the GBM underlying: (a) the option value at time $0$ is in the money where $M_0 = 2$, $F^C = 0.005$, $CTR=0.3$, $r=0.05$, $T = 31/365$ and $\sigma = 0.5$; and (b) the option value at time $0$ is out of the money where $M_0 = 2$, $F^C = 0.075$, $CTR=0.3$, $r=0.05$, $T = 31/365$ and $\sigma = 0.5$. Detailed descriptions of notations and terminology are provided in Table~\ref{tab:notation_summary}.}
\label{fig:lattice_convergence_comparison}
\end{figure*}

We follow a general economic settings and consider that the advertiser is risk-neutral so that he exercises the ad option only if the option payoff is maximized~\citep{Wilmott_2006_1}. We use the so-called risk-neutral probability measure for option pricing~\citep{Bjork_2009}. In finance, it is defined by the statement that the expected risky return of an asset is equal to a risk-less bank interest return. In the online advertising environment, the risk-neutral probability measure $\mathbb{Q} = (q, 1-q)$ satisfies the following equation 
\begin{equation}
\widetilde{r} M_0 \equiv q u M_0 + (1-q) d M_0,
\end{equation}
where $\widetilde{r} = (1 + \widehat{r})$ is the risk-less return over the period from time $0$ to time $1$, $u = M_1^{\{u\}}/M_0$ and $d = M_1^{\{d\}}/M_0$ are the movement scales of CPM. Therefore, we can obtain the risk-neutral transition probability $q = (\widetilde{r} - d)/(u - d)$. Note that here $q$ equals to $q_1$ in Table~\ref{tab:lattice_method_review}, which describes the probability that CPM moves upward in time $1$. Since the option value can be considered as a bivariate function of time and underlying price, the option value at time $0$ can be obtained by discounting the expected option value at time $1$ under $\mathbb{Q} = (q, 1-q)$~\citep[see Martingale]{Bjork_2009}. The option value at time $1$ is actually the option payoff; therefore, the option price at time $0$ can be obtained by discounting the expected payoff, that is  
\begin{align}
\pi_0 = & \ \widetilde{r} ^{-1} \mathbb{E}_{\mathbb{Q}}[\Phi_1] 
= \widetilde{r} ^{-1} \Big( q \Phi_1^{\{u\}} + (1-q) \Phi_1^{\{d\}} \Big). \label{eq:option_price_cpm_cpc_2_step} 
\end{align}

This option price $\pi_0$ is fair because it rules out arbitrage~\citep{Varian_1987,Bjork_2009}. Arbitrage means that an advertiser can obtain a profit larger or smaller than the risk-less bank interest rate with certainty. Consider if the option price is overestimated, i.e., $\pi_0 > \widetilde{r} ^{-1} ( q \Phi_1^{\{u\}} + (1-q) \Phi_1^{\{d\}} )$, the advertiser can sell short an ad option at time $0$ and save the money into bank to get the risk-less profit $\widetilde{r} \pi_0 - ( q \Phi_1^{\{u\}} + (1-q) \Phi_1^{\{d\}} )$. Converse strategies can be used to obtain arbitrage if the option price is underestimated. Up to this point, we have discussed the option pricing framework that is the one-step binomial method, initially proposed by~\citet*{Sharpe_1978}. For a multi-step binomial lattice, as shown in Figure~\ref{fig:lattice_framework}(a), the possible values of CPM and the corresponding risk-neutral transition probabilities can be estimated directly by investigating various combinations of each one-step model, so the option price $\pi_0$ can be obtained as follows   
\begin{align}
\pi_0 
= & \ \widetilde{r}^{-n} \ 
\sum_{j = 0}^n 
{n \choose j}
q^j (1-q)^{n-j} 
\bigg(\frac{u^j d^{n-j} M_0}{1000 H}  - F^C \bigg)^+. \label{eq:option_price_cpm_cpc_crr_1} 
\end{align}

If for any $j \geq j^*$, $u^j d^{n-j} M_0 / (1000 H) \geq F^C$, then
\begin{align}
\pi_0 
= & \  
\frac{M_0}{1000 H}   
\sum_{j = j^*}^n 
{n \choose j} 
\widetilde{q}^j 
(1-\widetilde{q})^{n-j} 
 -  F^C \widetilde{r}^{-n} 
\sum_{j = j^*}^n 
{n \choose j}
q^j (1-q)^{n-j} \nonumber \\
= & \  \frac{M_0}{1000 H} 
\psi(j^*, n, \widetilde{q})
-  F^C
\widetilde{r}^{-n} \psi(j^*, n, q),
\label{eq:option_price_cpm_cpc_crr} 
\end{align} 
where $\widetilde{q} = q \times (u/\widetilde{r})$. If each time step $\Delta t = T/n$ is sufficiently small, a continuous-time closed-form formula for $\pi_0$ can be obtained as follows 
\begin{align}
\pi_0 = & \ 
\frac{M_0}{1000 H} \mathcal{N}( \varsigma_1) - F^C e^{- r T} \mathcal{N} (\varsigma_2),\\
\varsigma_ 1 = & \ 
\frac{1}{\sigma \sqrt{T}} \bigg( 
\ln \big\{\frac{M_0}{1000 H F^C} \big\} + (r + \frac{1}{2}\sigma^2 ) T 
\bigg), \\
\varsigma_ 2 = & \ 
\varsigma_ 1 - \sigma \sqrt{T},
\label{eq:option_price_cpm_cpc_bsm}
\end{align}
which is very similar to the BSM model~\citep{Black_1973,Merton_1973}.

Figure~\ref{fig:lattice_framework}(b) exhibits a trinomial lattice. There are 6 parameters: $u, m, d$ are state movement scales; $q_1, q_2, q_3$ are the corresponding risk-neutral transition probabilities. These parameters uniquely determine the movement of CPM, which then determines a unique value of an ad option written on CPM. They must be restricted such that the constructed trinomial lattice converges to the log-normal distribution of CPM in continuous time (i.e., the GBM assumption). We use the \emph{moment matching technique}~\citep{Cox_1979} to define the basic restrictions as follows:
\begin{align}
q_1 + q_2 + q_3 = & \ 1,\\
q_1 u + q_2 m + q_3 d = & \ \gamma = e^{ r \Delta t},\\
q_1 u^2 + q_2 m^2 + q_3 d^2 = & \ \gamma^2 \zeta = e^{2 r \Delta t} e^{\sigma^2 \Delta t}
\end{align}
where $0 \leq q_1, q_2, q_3 \leq 1$. Since there are 6 parameters, 3 additional equations are necessary to define a unique solution. Here we examine the additional conditions discussed by previous research~\citep{Boyle_1988,Kamrad_1991,Tian_1993} and use the same settings to price a display ad option. 

Figure~\ref{fig:lattice_convergence_comparison} compares the convergence performance of discussed binomial and trinomial lattice methods for option pricing. Eq.~(\ref{eq:option_price_cpm_cpc_bsm}) is used as the golden line to examine how quickly that the calculated option price from lattice methods approximate to the closed-form value (because these methods are all based on the GBM assumption). Figure~\ref{fig:lattice_convergence_comparison}(a) illustrates the situation when the option value at time $0$ is in the money (i.e., $M_0/(1000 H) \geq F^C$) and Figure~\ref{fig:lattice_convergence_comparison}(b) shows the out of the money case~(i.e., $M_0/(1000 H) < F^C$). Several findings are worth mentioning here. First, the convergence rate of the trinomial lattice is fast than that of the binomial lattice; however, more nodes need to be computed for the former, i.e., $(n+1)^2$ nodes for the trinomial lattice while there are only $(n+1)(n+2)/2$ nodes for binomial lattice. Second, the Tian-TRIN~\citep{Tian_1993} model has a better convergence performance than the others.

\section{Censored Binomial Lattice for the SV Underlying Model}
\label{ao:censored_binomial_lattice}

When the GBM assumption is not valid empirically, the SV model can be used to describe the underlying price movement. Let us extend the case whereby an ad option allows its buyer to pay a fixed CPC for display impressions. The SV model for the uncertain winning payment CPM can be expressed as follows:
\begin{align}
d M (t) = & \ \mu M(t) dt + \sigma(t) M(t) d W(t), \label{eq:model_sde_price}\\
d \sigma(t) = & \ \kappa (\theta - \sigma(t)) dt + \delta \sqrt{\sigma(t)} d Z(t),
\label{eq:model_sde_volatility}
\end{align}
where $W(t)$ and $Z(t)$ are standard Brownian motions under the real world probability measure $\mathbb{P}$ satisfying $\mathbb{E}[d W(t)  dZ(t)] = 0$, and $\mu$ and $\sigma(t)$ are the constant drift and volatility of CPM, and $\kappa$, $\theta$, $\delta$ are the volatility parameters. The drift factor $\kappa (\theta - \sigma(t))$ ensures the mean reversion of $\sigma(t)$ towards its long-term value $\theta$. The volatility factor $\delta \sqrt{\sigma(t)}$ avoids the possibility of negative $\sigma(t)$ for all positive values of $\kappa$ and $\theta$. It is worth noting that the proposed model is very similar to the Heston model~\citep{Heston_1993} while the significant difference is that the hidden layer is driven by $d \sigma(t)$ rather than $d \sigma(t)^2$. Let $X(t) = \ln(M(t))$, Eq.~(\ref{eq:model_sde_price}) can be re-written as the following risk-neutral form:
\begin{align}\label{eq:sv_no_arbitrage}
d X(t) = & \ \bigg( r - \frac{ \sigma^2(t)}{2} \bigg) dt + \sigma(t) d W^{\mathbb{Q}}(t), 
\end{align}
where $r$ is the constant continuous-time risk-less interest rate and $W^{\mathbb{Q}} := W(t) + \int_0^t \frac{\mu - r}{\sigma(s)} d s$ is a standard Brownian motion under the risk-neutral probability measure $\mathbb{Q}$, so $\mathbb{E}[d W^{\mathbb{Q}}(t) dZ(t)] = 0$. The process $X(t)$ can be weakly approximated by a series of binomial processes, say $\widetilde{X}(t_i), i = 1, \ldots, n$. For more details about the approximation conditions, see~\citep*{Nelson_1990}. We will briefly verify these conditions in the following discussion.

\begin{figure*}[t]
\centering
\includegraphics[width=0.85\linewidth]{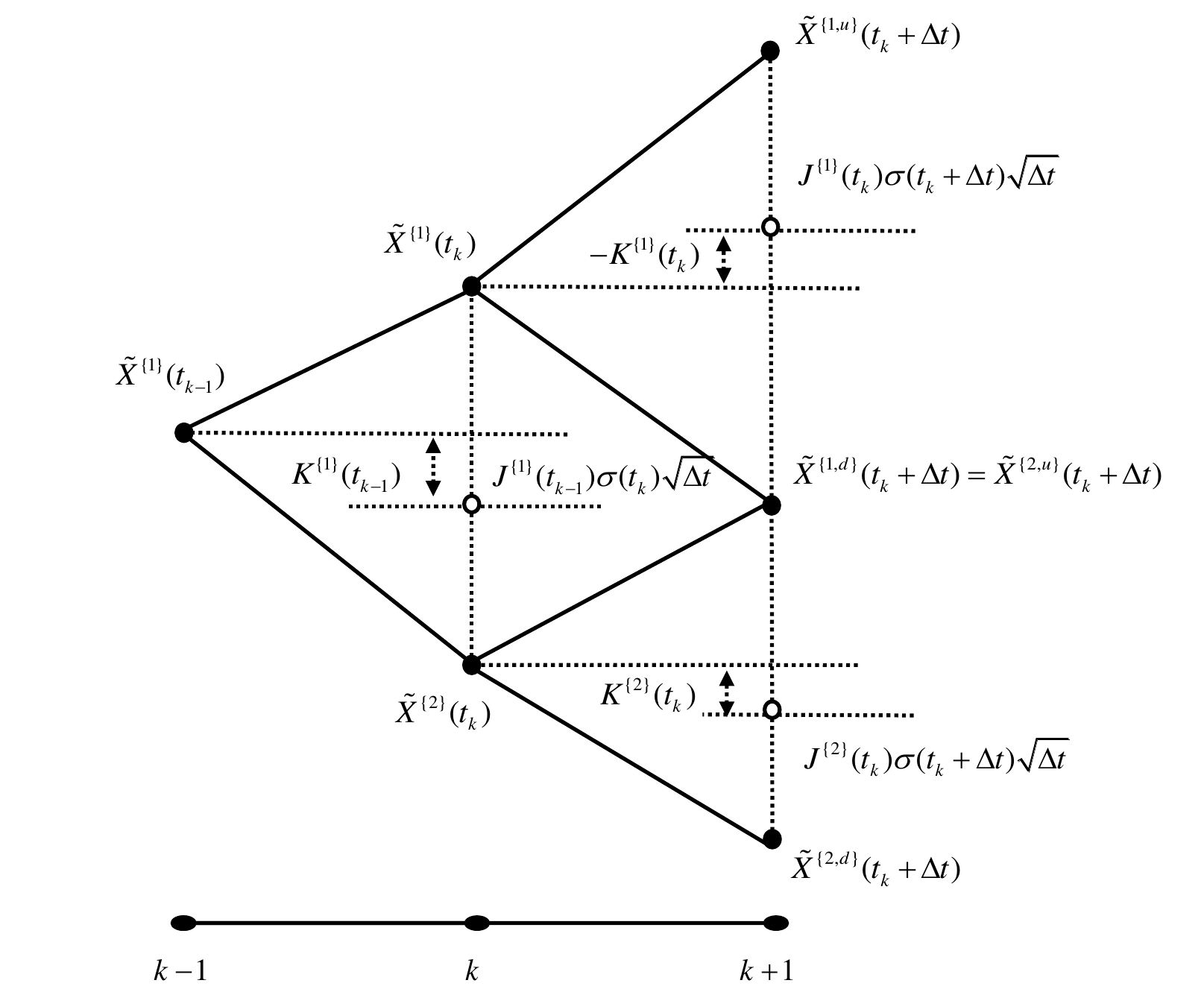}
\caption{Censored binomial lattice for the SV underlying. Detailed description of notations is provided in Table~\ref{tab:notation_summary}.}
\label{fig:lattice_binomial_nongbm}
\end{figure*}

In Algorithm~\ref{algo:cpm_cpc_censored_binomial}, we present our method of calculating the option price for a display ad option whose underlying is the SV model. Simply, a binomial lattice for $\widetilde{X}(t_i)$ is first constructed to approximates $X(t)$ weakly. The lattice is constructed from time step 0 to time step $n$, and at each time step, nodes are calculated from top to bottom. In the following discussion, the mathematical details of Steps~\ding{172}-\ding{174} are introduced.

\textbf{Step~\ding{172}} We start from the first node $\widetilde{X}^{\{1\}}(t_k)$ in Figure~\ref{fig:lattice_binomial_nongbm}, whose two successors can be expressed as follows
\begin{align}
\widetilde{X}^{\{1,u\}}(t_k +\Delta t) = & \ (J^{\{1\}}(t_k)+1) \sigma(t_k +\Delta t) \sqrt{\Delta t} \nonumber \\ 
& + \bigg( r - \frac{ \sigma^2(t_k+\Delta t)}{2} \bigg) \Delta t, \label{eq:sv_u} \\
\widetilde{X}^{\{1,d\}}(t_k +\Delta t) = & \ (J^{\{1\}}(t_k)-1) \sigma(t_k +\Delta t) \sqrt{\Delta t} \nonumber \\
& + \bigg( r - \frac{ \sigma^2(t_k+\Delta t)}{2} \bigg) \Delta t,
\label{eq:sv_d}
\end{align}
where $J^{\{1\}}(t_k) \sigma(t_k +\Delta t) \sqrt{\Delta t}$ is the point on the grid closest to $\widetilde{X}^{\{1\}}(t_k)$, given by
\begin{align}
J^{\{1\}}(t_k) = & \ \inf_{J^* \in \mathbb{N}} \ \Big\vert \ J^* \sigma (t_k+ \Delta t) \sqrt{\Delta t} - \widetilde{X}^{\{1\}} (t_k) \ \Big\vert.
\label{eq:j_node_1}
\end{align}

Eqs.~(\ref{eq:sv_u})-(\ref{eq:sv_d}) can be rewritten in terms of their conditional increments:
\begin{align}
 & \ \widetilde{X}^{\{1,u\}}(t_k +\Delta t) - \widetilde{X}^{\{1\}}(t_k) \nonumber \\
= & \ \sigma(t_k +\Delta t) \sqrt{\Delta t} - K^{\{1\}}(t_k) + \bigg( r - \frac{ \sigma^2(t_k+\Delta t)}{2} \bigg) \Delta t, \label{eq:sv_increment_u} \\
& \ \widetilde{X}^{\{1,d\}}(t_k +\Delta t) - \widetilde{X}^{\{1\}}(t_k) \nonumber \\
= & \ - \sigma(t_k +\Delta t) \sqrt{\Delta t} - K^{\{1\}}(t_k) + \bigg( r - \frac{ \sigma^2(t_k+\Delta t)}{2} \bigg) \Delta t,
\label{eq:sv_increment_d}
\end{align}
where $K^{\{1\}}(t_k)$ is the grid adjusting parameter for the successors of the first node at time $t_k$. As shown in Figure~\ref{fig:lattice_binomial_nongbm}, the value of $K^{\{i\}}(t_k)$, $i=1,2,\ldots, k$, can be either positive or negative, To satisfy the approximation condition $\lim_{\Delta t \rightarrow 0} | X(t_k +\Delta t) - X(t_k) | = 0$, the following equation holds:
\begin{align}
\mathbb{E} \Big[ \widetilde{X}^{\{1\}}(t_k +\Delta t) - \widetilde{X}^{\{1\}}(t_k) \mid \mathcal{F}(t_k) \Big] 
= & \
\bigg( r - \frac{ \sigma^2(t_k +\Delta t)}{2} \bigg) \Delta t.
\end{align}
Then, we can obtain a system of equations
\begin{align}
\Big( \sigma(t_k +\Delta t) \sqrt{\Delta t} - K^{\{1\}}(t_k) \Big) \frac{q_1^{\{1\}}(t_k)}{Q^{\{1\}}(t_k)} \ + \ & \nonumber \\
\Big( - \sigma(t_k+\Delta t) \sqrt{\Delta t} - K^{\{1\}}(t_k) \Big)\frac{q^{\{1\}}_2(t)}{Q^{\{1\}}(t)} \ & = 0, \nonumber \\ 
q_1^{\{1\}}(t_k) + q_2^{\{1\}}(t_k) \ & = Q^{\{1\}}(t),\nonumber 
\end{align}
where $q_1^{\{1\}}(t_k)$ and $q_2^{\{1\}}(t_k)$ are the risk-neutral probabilities that the successor of the first node at time $t_k$ rises or falls in time $t_k + \Delta t$, and $Q^{\{1\}}(t_k)$ is the risk-neutral probability for the first node at time $t_k$. Solving the above equations then gives
\begin{align}
q_1^{\{1\}}(t_k)
= & \
\begin{cases}
\frac{Q^{\{1\}}(t_k)}{2} \ \bigg(1 + \frac{K^{\{1\}}(t_k)}{\sigma(t_k +\Delta t) \sqrt{\Delta t}} \bigg),
&  \\
\ \ \ \ \ \ \textrm{if } 0 \leq \frac{Q^{\{1\}}(t_k)}{2} \ \bigg(1 + \frac{K^{\{1\}}(t_k)}{\sigma(t_k +\Delta t) \sqrt{\Delta t}} \bigg) \leq  Q^{\{1\}}(t_k), \\
0, \ \ \ \ \ \ \ \ \ \ \
\textrm{if } \frac{Q^{\{1\}}(t_k)}{2} \ \bigg(1 + \frac{K^{\{1\}}(t_k)}{\sigma(t_k +\Delta t) \sqrt{\Delta t}} \bigg) < 0, \\
Q^{\{1\}}(t_k), \ \textrm{if } \frac{Q^{\{1\}}(t_k)}{2} \ \bigg(1 + \frac{K^{\{1\}}(t_k)}{\sigma(t_k +\Delta t) \sqrt{\Delta t}} \bigg) \geq Q^{\{1\}}(t_k), 
\\
\end{cases}
\nonumber \\
= & \ \Bigg(
Q^{\{1\}}(t_k) \wedge 
\frac{Q^{\{1\}}(t_k)}{2} \bigg(1 + \frac{K^{\{1\}}(t_k)}{\sigma(t_k +\Delta t) \sqrt{\Delta t}} \bigg)
\Bigg)^{\hspace{-2pt} +}, 
\label{eq:censored_prob_u}\\
q_2^{\{1\}}(t_k)
= & \ Q^{\{1\}}(t_k) - q_1^{\{1\}}(t_k). \label{eq:censored_prob_d}
\end{align}

\begin{algorithm}[tp]
\caption{Censored binomial lattice method for pricing a display ad option with the SV underlying model. Detailed description of notations is provided in Table~\ref{tab:notation_summary}.}
\label{algo:cpm_cpc_censored_binomial}
\begin{algorithmic}
\Function{\texttt{OptionPricingCBL}}{$M_0, \sigma_0, \kappa, \theta, \delta, H, T, n, r, F^C$}
	\State $\Delta t \leftarrow T/n$; $\widetilde{r} \leftarrow e^{r \Delta t}$;
	\For{$\; k \leftarrow 0$ to $n-1$}
		\For{$i \in $ nodes in time step $k$}
		\If{$i=1$}		
			\State Step~\ding{172};
		\Else			  
			\State Step~\ding{173};	    
		\EndIf
		\EndFor	
	\EndFor
	\State $\pi_0 \leftarrow$~Eq.~(\ref{eq:option_price_censored_binomial_lattice}) (see Step~\ding{174});
\EndFunction
\end{algorithmic}
\end{algorithm}

Eqs.~(\ref{eq:censored_prob_u}) and~(\ref{eq:censored_prob_d}) show that transition probabilities $q_1^{\{1\}}(t_k)$ and $q_2^{\{1\}}(t_k)$ are censored in the approximation. 

\textbf{Step~\ding{173}} The successors of other nodes can be constructed in the same manner as that of $\widetilde{X}^{\{1\}}(t_k) $. Since the transition probabilities are censored directly at each node, $K^{\{i\}}(t_k)$, $J^{\{i\}}(t_k)$ and $Q^{\{i\}}(t_k)$ can be calculated sequentially from top to bottom alongside the lattice construction for the underlying price. The nodes need to be kept the recombining pattern; therefore, the following equations hold for $1 \leq i \leq k$: 
\begin{align}
  & \ \widetilde{X}^{\{i,d\}}(t_k +\Delta t) \nonumber \\
= & \ (J^{\{i\}}(t_k)-1) \sigma(t_k +\Delta t) \sqrt{\Delta t} 
+ \bigg( r - \frac{ \sigma^2(t_k+\Delta t)}{2} \bigg) \Delta t \nonumber \\
= & \ \widetilde{X}^{\{i+1,u\}}(t_k +\Delta t) \nonumber \\
= & \ (J^{\{i+1\}}(t_k)+1) \sigma(t_k +\Delta t) \sqrt{\Delta t} 
+ \bigg( r - \frac{ \sigma^2(t_k+\Delta t)}{2} \bigg) \Delta t, \nonumber
\end{align}
therefore 
\begin{align*}
J^{\{i+1\}}(t_k) = & \ J^{\{i\}}(t_k) - 2, \\
K^{\{i+1\}}(t_k) = & \ J^{\{i+1\}}(t_k) \sigma(t_k +\Delta t) \sqrt{\Delta t} - \widetilde{X}^{\{i+1\}}(t_k). 
\end{align*}
The transition probabilities for the node $\widetilde{X}^{\{i+1\}}(t_k)$ can be then estimated by Eqs.~(\ref{eq:censored_prob_u})-(\ref{eq:censored_prob_d}). Hence, the rolling risk-neutral probability distribution $Q^{\{i\}}(t_k)$ for each node can be quickly computed as follows:
\[
Q^{\{i\}}(t_k+\Delta t) = 
\left\{
\begin{array}{ll}
q_1^{\{1\}}(t_k), & \textrm{if } i = 1, \\
q_2^{\{i-1\}}(t_k) + q_1^{\{i\}}(t_k), & \textrm{if } 1 < i < k+1, \\
q_2^{\{k+1\}}(t_k), & \textrm{if } i = k+1, \\
\end{array}
\right.
\]
subjected to the initial condition $Q(t_0) = 1$. 

\textbf{Step~\ding{174}} The binomial lattice can be constructed by steps~\ding{172}-\ding{173} for each time step until the contract expiration date. Finally, the option price can be obtained as follows:
\begin{equation}
\pi_0 = \widetilde{r}^{-n} \sum_{i=1}^{n+1} Q^{\{i\}}(t_n) 
\bigg(\frac{1}{1000 H} e^{\widetilde{X}^{\{i\}}(t_n)} - F^C \bigg)^+.
\label{eq:option_price_censored_binomial_lattice}
\end{equation}
Similar to Eq.~(\ref{eq:option_price_cpm_cpc_crr}), Eq.~(\ref{eq:option_price_censored_binomial_lattice}) is also the discrete form of the risk-neutral terminal pricing~\citep{Bjork_2009}. 

\begin{figure*}[t]
\centering
\includegraphics[width=0.825\linewidth]{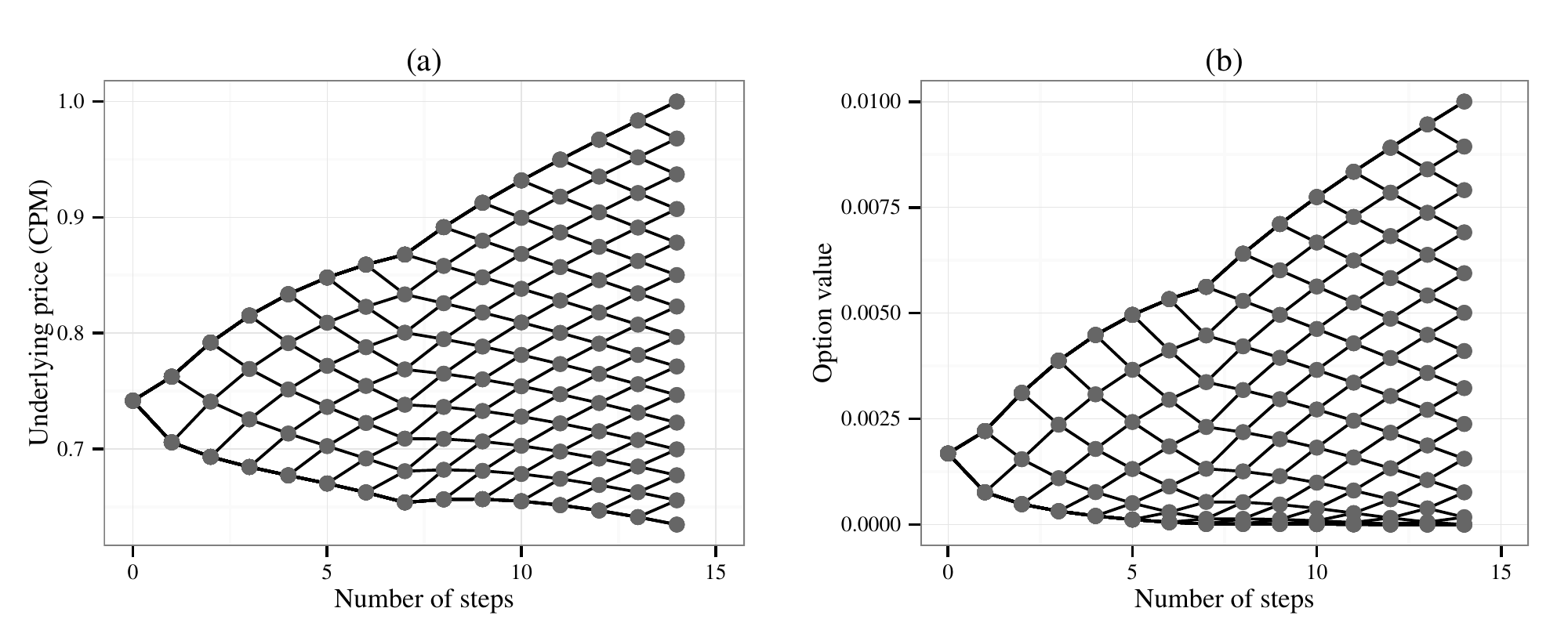}
\caption{Empirical example of binomial lattices for an ad slot from the SSP dataset: (a) the censored binomial lattice for CPM based on the SV model, where $r=0.05, T=0.0384, n = 14, CPM=0.7417, \sigma_0 = 0.8723, \kappa = 96.4953, \theta = 0.2959, \delta = 14.9874$; (b) the censored binomial lattice for the option value. The model parameters are estimated based on the training data.}
\label{fig:nongbm_lattice_example_1}
\includegraphics[width=0.825\linewidth]{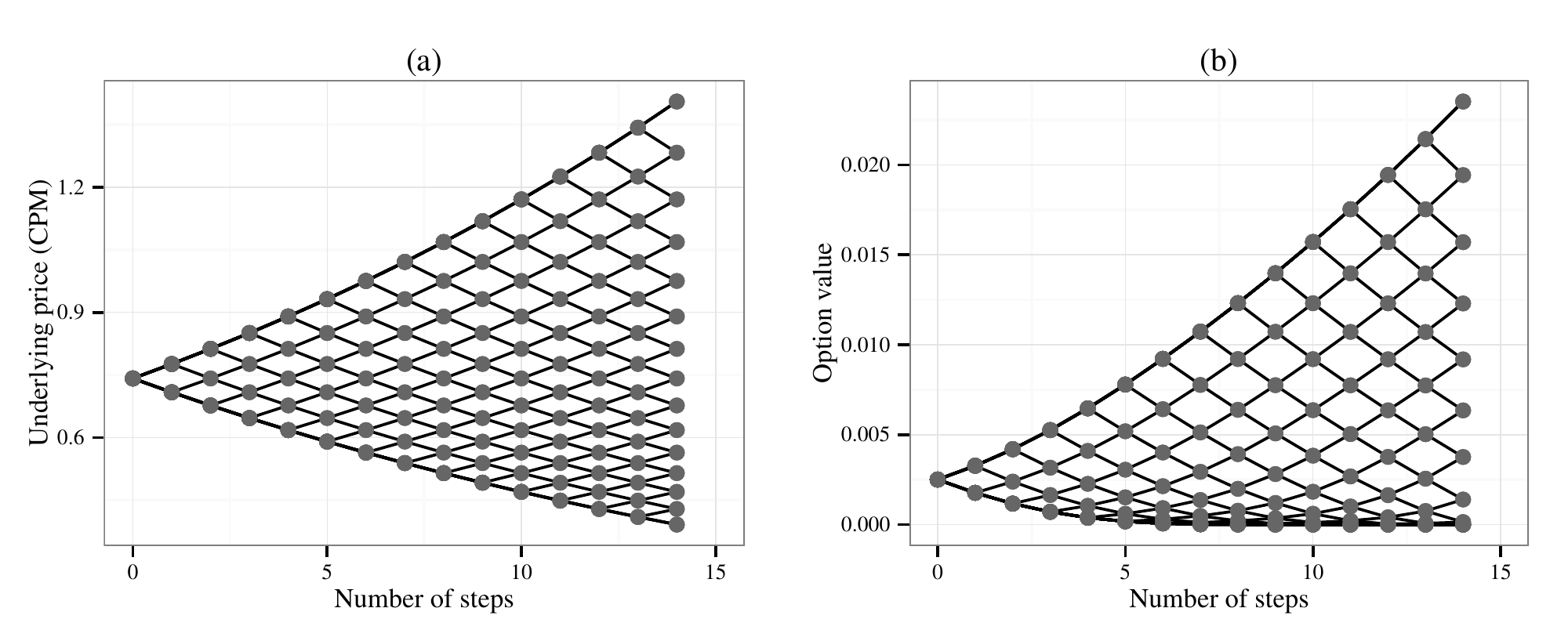}
\caption{Example of binomial lattices for the same ad slot in Figure~\ref{fig:nongbm_lattice_example_1}: (a) the CRR binomial lattice for CPM based on the GBM model, where $r=0.05, T=0.0384, n = 14, CPM=0.7417, \sigma_0 = 0.8723$. Here we use the same parameters' values in Figure~\ref{fig:nongbm_lattice_example_1}; (b) the CRR binomial lattice for the option value.}
\label{fig:nongbm_lattice_example_2}
\end{figure*}

In the above discussion, we actually followed~\citet*{Florescu_2005} to construct the binomial lattice and used variables $K^{\{i\}}(t_k)$ and $J^{\{i\}}(t_k)$ to tune the grid so that the constructed binomial framework is recombining. Compared to \citet*{Florescu_2005}, our method simplifies the lattice construction process by censoring the probabilities at each node directly. In the meantime, the structure satisfies the approximation conditions proposed by~\citet*{Nelson_1990}. Figure~\ref{fig:nongbm_lattice_example_1} presents an empirical example of constructing a censored binomial lattice for pricing a display ad option written on an ad slot from a SSP in the UK. The given values of the model parameters are estimated from the training data. Figure~\ref{fig:nongbm_lattice_example_1}(a) shows a censored binomial lattice for the underlying CPM and Figure~\ref{fig:nongbm_lattice_example_1}(b) illustrates how the option value is calculated backward iteratively from the expiration date to time $0$. For the sake of comparison, Figure~\ref{fig:nongbm_lattice_example_2} illustrates the binomial lattices constructed by the CRR model with the same parameter settings. Obviously, the changing volatility can be found in Figure~\ref{fig:nongbm_lattice_example_1}(a) while~\ref{fig:nongbm_lattice_example_2}(a) exhibits a constant volatility over time. We find that the option price given by the SV model is slightly smaller than that of the CRR model. This is because the long-term mean value of volatility is $0.2959$, smaller than its initial value $0.8723$. Therefore, the drift drags the volatility downside to its long-term level and the option value based on the SV model contains less risk than the CRR model.

\section{Empirical Evaluation}\label{ao:experiments}

This section presents our experimental results. We examine the GBM assumption with real advertising data, compare the fitness of underlying models, validate the proposed lattice method via Monte Carlo simulations, analyse if an advertiser can have better deliveries under a fixed daily budget, and discuss the effects on the publisher's (or search engine's) revenue.

\subsection{Datasets and Experimental Design}

\begin{table*}[t]
\centering
\caption{Summary of datasets for experiments.}
\label{tab:datasets}
\begin{tabular}{r|r|r}
\hline
Dataset & SSP & Google AdWords\\
\hline
Period     & 08/01/2013 - 14/02/2013 & 26/11/2011 - 14/01/2013\\
Number of ad slots or keywords  & 31 	& 557  \\
Number of advertisers   & 374 		& $\times$\\ 
Number of impressions   & 6646643 	& $\times$\\ 
Number of bids          & 33043127   	&  $\times$ \\
Winning payment price           & $\surd$   	&  $\surd$ \\
Bid quote       & GBP/CPM  		& GBP/CPC\\ 
\hline
\end{tabular}
\vspace{5pt}
\caption{Experimental settings of the SSP dataset. }
\label{tab:exp_setting_ssp}
\begin{tabular}{r|r}
\hline
Training set (31 days) & Development \& test set (7 days)\\
\hline
08/01/2013-07/02/2013 & 08/02/2013-14/02/2013\\
\hline
\end{tabular}
\vspace{5pt}
\caption{Experimental settings of the Google AdWords dataset.}
\label{tab:exp_setting_google} 
\begin{tabular}{r|r|r|r}
\hline
Market &  Group & Training set (31 days) & Development~\&~test set (31 days) \\
\hline
\multirow{4}{*}{US}  & 1 & 25/01/2012-24/02/2012 & 24/02/2012-25/03/2012 \\
                     & 2 & 30/03/2012-29/04/2012 & 29/04/2012-31/05/2012 \\
                     & 3 & 10/06/2012-12/07/2012 & 12/07/2012-17/08/2012 \\
                     & 4 & 10/11/2012-11/12/2012 & 11/12/2012-10/01/2013 \\                    
\hline
\multirow{4}{*}{UK}  & 1 & 25/01/2012-24/02/2012 & 24/02/2012-25/03/2012 \\
                     & 2 & 30/03/2012-29/04/2012 & 29/04/2012-31/05/2012 \\
                     & 3 & 12/06/2012-13/07/2012 & 13/07/2012-19/08/2012 \\
                     & 4 & 18/10/2012-22/11/2012 & 22/11/2012-24/12/2012 \\                    
\hline
\end{tabular}
\end{table*}

\begin{figure*}[t]
\centering
\includegraphics[width=0.825\linewidth]{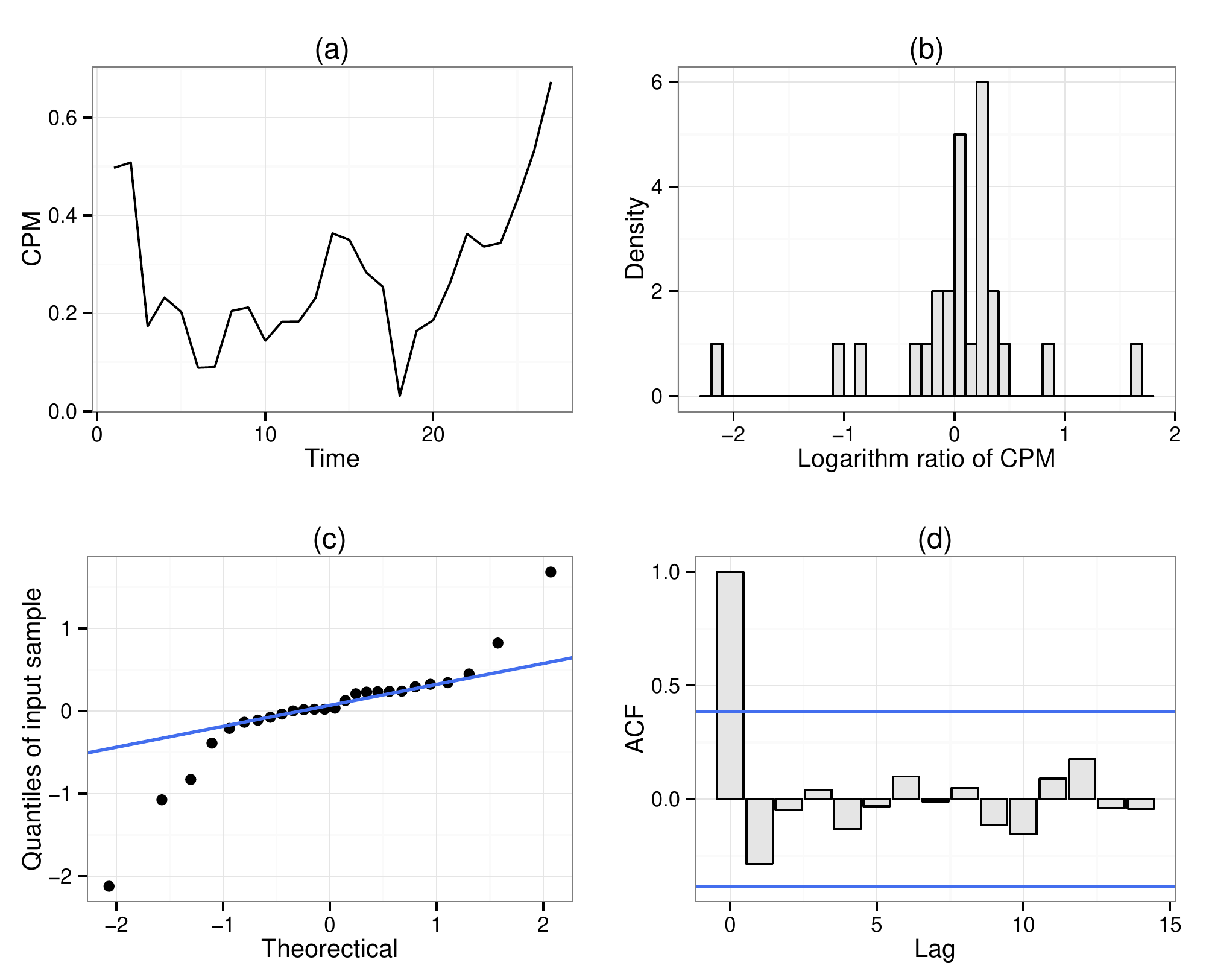} 
\caption{Empirical example of testing the GBM conditions of an ad slot from the SSP dataset: (a) the plot of the average daily winning payment CPMs from auctions; (b) the histogram of the logarithm ratios of the CPM, i.e., $\ln(M_{i+1}/M_i)$, $i = 1, \ldots, n-1$; (c) the QQ plot of the logarithm ratios; (d) the plot of the ACFs of the logarithm ratios. The Shapiro-Wilk test is with p-value 0.0009 and the Ljung-Box test is with p-value 0.1225.}
\label{fig:gbm_test_display_ads_example}
\end{figure*}

\begin{figure*}[tp]
\centering
\includegraphics[width=0.825\linewidth]{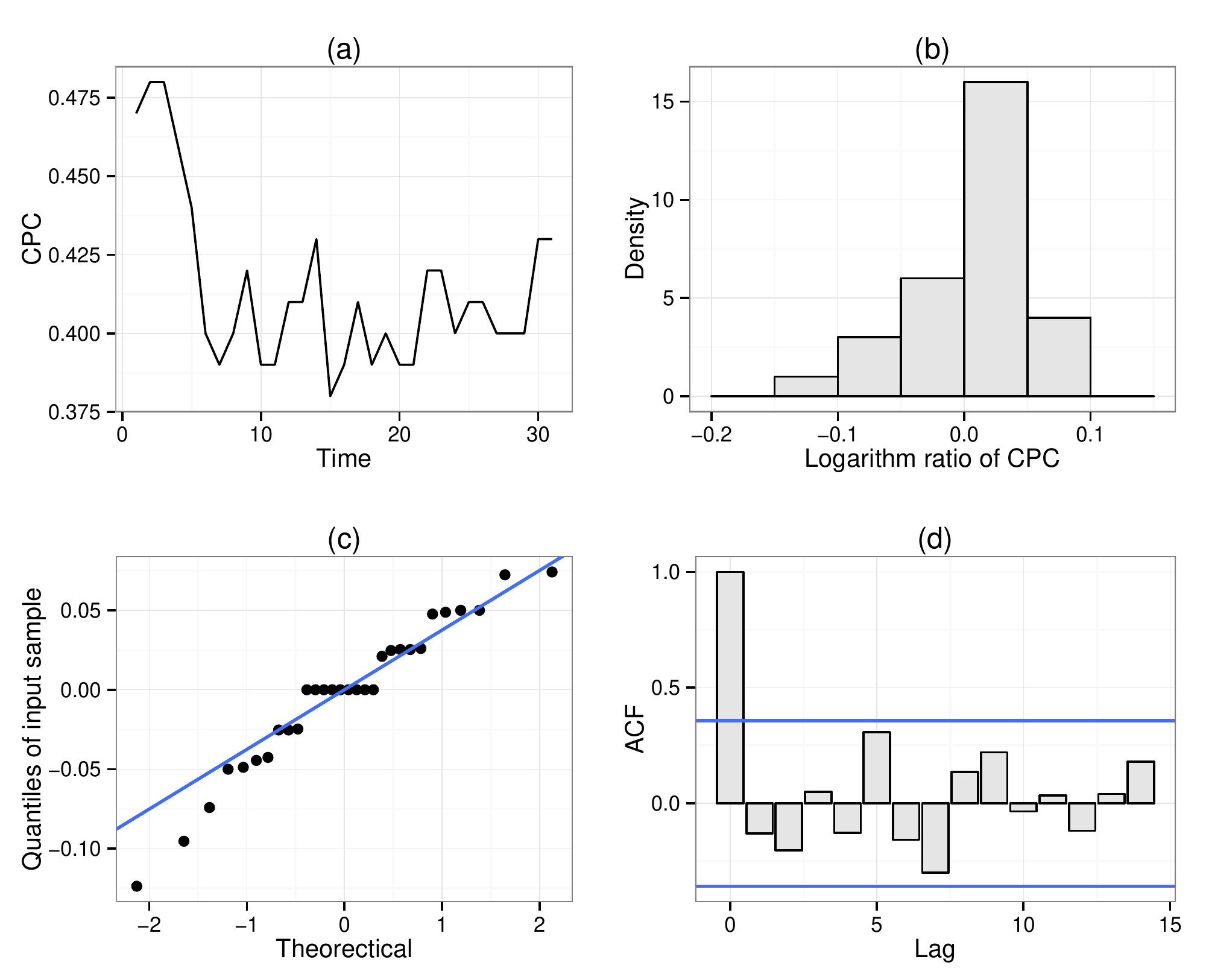}
\caption{Empirical example of testing the GBM conditions of the keyword \lq\lq{}canon 5d\rq\rq{} from the Google AdWords dataset: (a) the plot of average daily winning payment CPCs; (b) the histogram of logarithm ratios of CPC, i.e., $\ln(C_{i+1}/C_i)$, $i = 1, \ldots, n-1$; (c) the QQ plot of the logarithm ratios; (d) the plot of the ACFs of the logarithm ratios. The Shapiro-Wilk test is with p-value 0.2144 and the Ljung-Box test is with p-value 0.6971.}
\label{fig:gbm_test_search_example}
\includegraphics[width=0.825\linewidth]{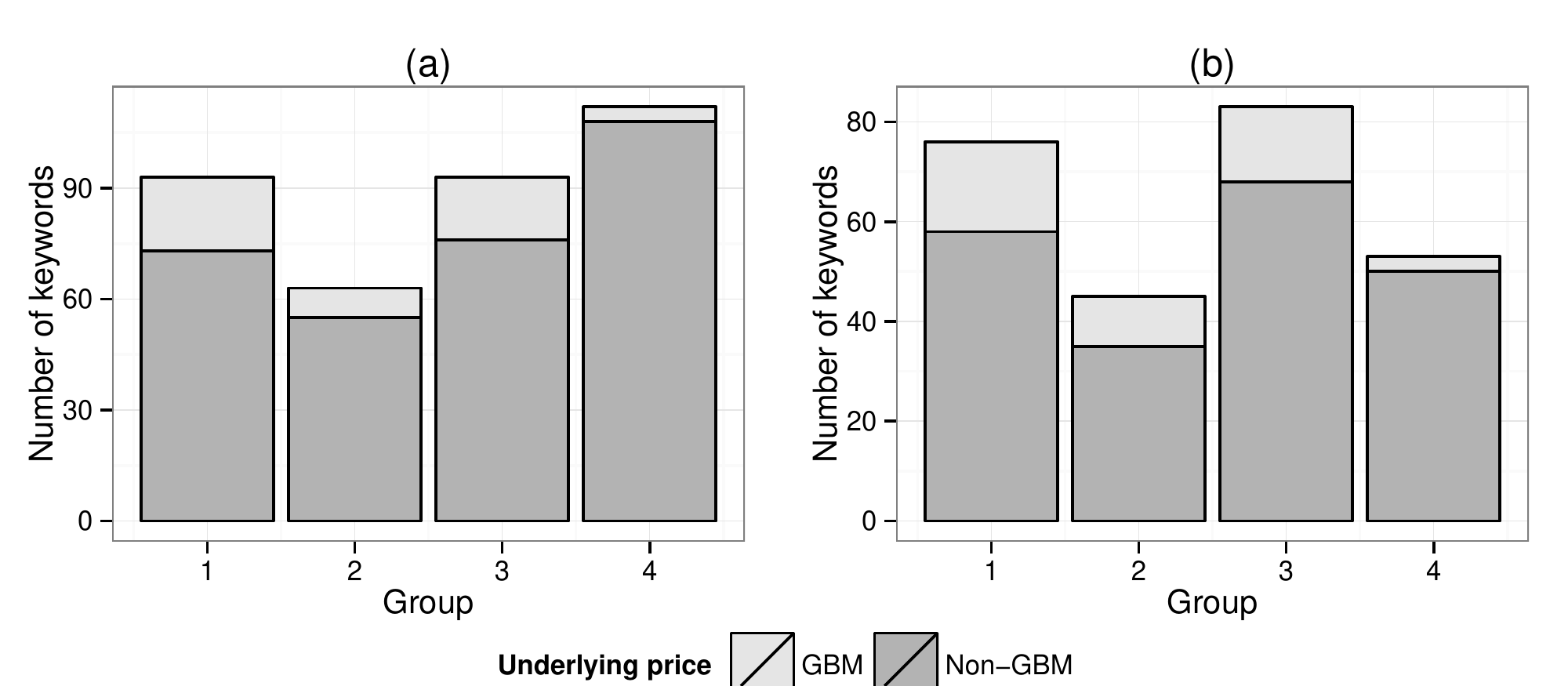}
\caption{Summary of the GBM conditions test for all keywords in the Google AdWords dataset.}
\label{fig:gbm_test_search_overview}
\end{figure*} 

Table~\ref{tab:datasets} presents the two datasets used in experiments: a RTB dataset from a SSP in the UK; and a sponsored search dataset from Google AdWords. The RTB dataset contains all advertisers' bids and the corresponding winning payment CPMs (per transaction). The Google dataset is obtained by using Google's Traffic Estimation service~\citep{Yuan_2012}. Tables~\ref{tab:exp_setting_ssp}-\ref{tab:exp_setting_google} illustrates our experimental settings. Each dataset is divided into several experimental groups and each group is specified with one training, one development and one test set. The model parameters are estimated in the training set. Display ad options are priced in the development set. The actual bids in the test set are used to examine the priced options. The default value of CTR is set to be $0.03$. 

\subsection{Fitness of GBM and SV Models}

The following two conditions hold if the GBM assumption is valid empirically: (i) the normality of the logarithm ratios of the winning payment price\footnote{The logarithm ratio of winning payment price $L_i$ is defined by $L_i = \ln(M_{i+1}/M_i)$ or $L_i = \ln(C_{i+1}/C_i)$.}; and (ii) the independence of the logarithm ratios from the previous data. Normality can be graphically checked by a histogram or Q-Q plot, and be statistically verified by the Shapiro-Wilk test~\citep{Shapiro_1965}; independence can be tested by the autocorrelation function (ACF)~\citep{Tsay_2005} and the Ljung-Box statistic~\citep{Ljung_1978}. It is worth noting that the above two conditions are necessary conditions while we follow~\cite{Marathe_2005} and consider the GBM assumption is valid empirically if they are not rejected by real data.  

Figure~\ref{fig:gbm_test_display_ads_example} presents an empirical example of testing the GBM assumption for an ad slot from the SSP dataset, where the underlying winning CPM cannot be described accurately as a GBM. In fact, none of the 31 ad slots in the SSP dataset satisfy the GBM model. Therefore, we use the SV model for the ad slots in the SSP dataset. Figure~\ref{fig:gbm_test_search_example} presents an example of a keyword from the Google dataset. The keyword's winning CPC satisfies the GBM assumption. The log-normality of CPC is validated in Figure~\ref{fig:gbm_test_search_example}(a)-(c) and the independence is confirmed by Figure~\ref{fig:gbm_test_search_example}(d). The overview results of the Google dataset is shown in Figure~\ref{fig:gbm_test_search_overview}. There are 14.25\% and 17.20\% of the keywords in the US and UK markets respectively that can be accurately described by the GBM model. We will price the remaining keywords using the SV model.

\begin{figure*}[tp]
\centering
\includegraphics[width=0.825\linewidth]{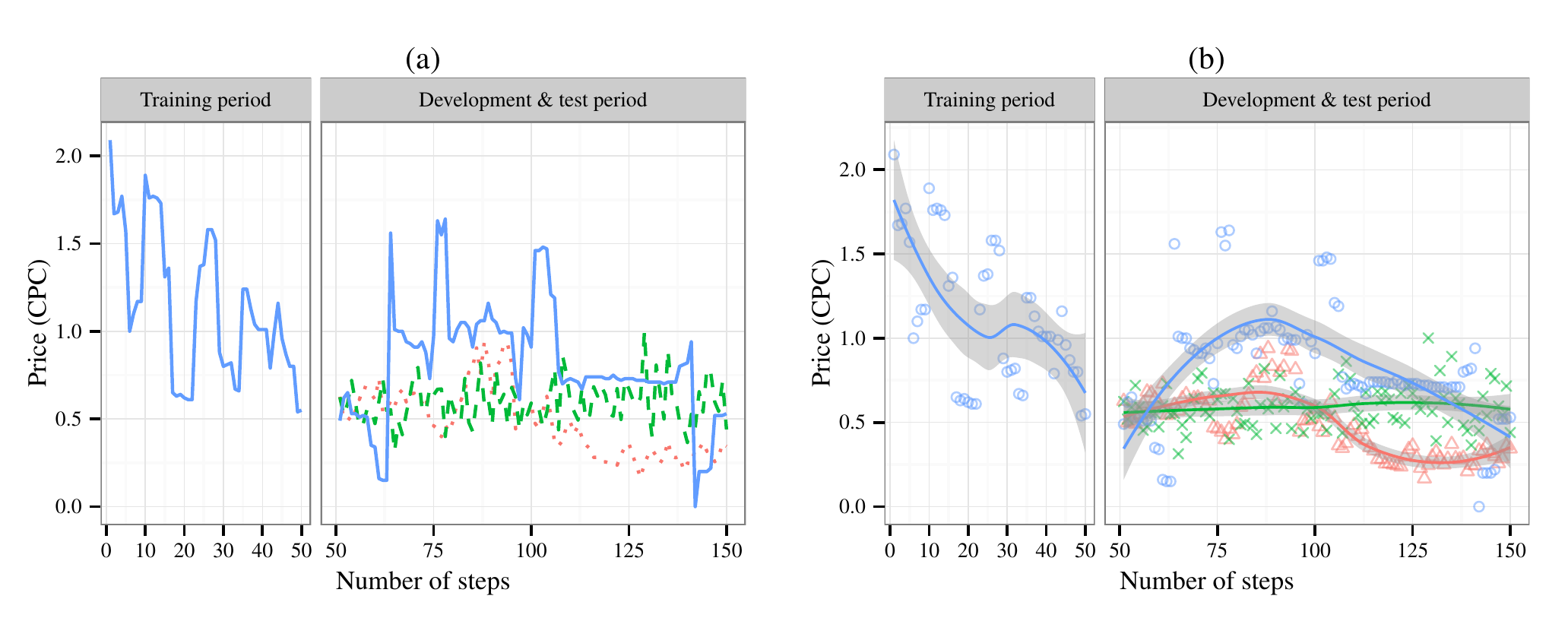}
\includegraphics[width=0.825\linewidth]{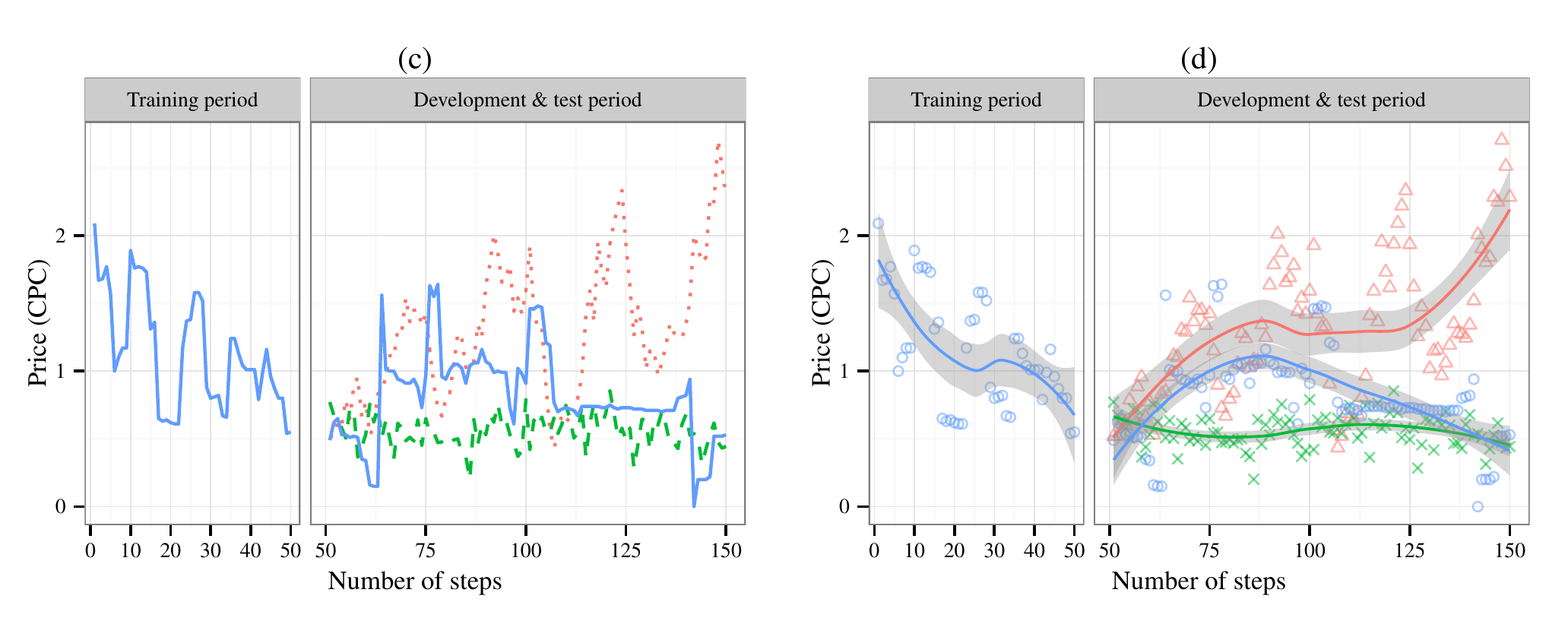}
\includegraphics[width=0.825\linewidth]{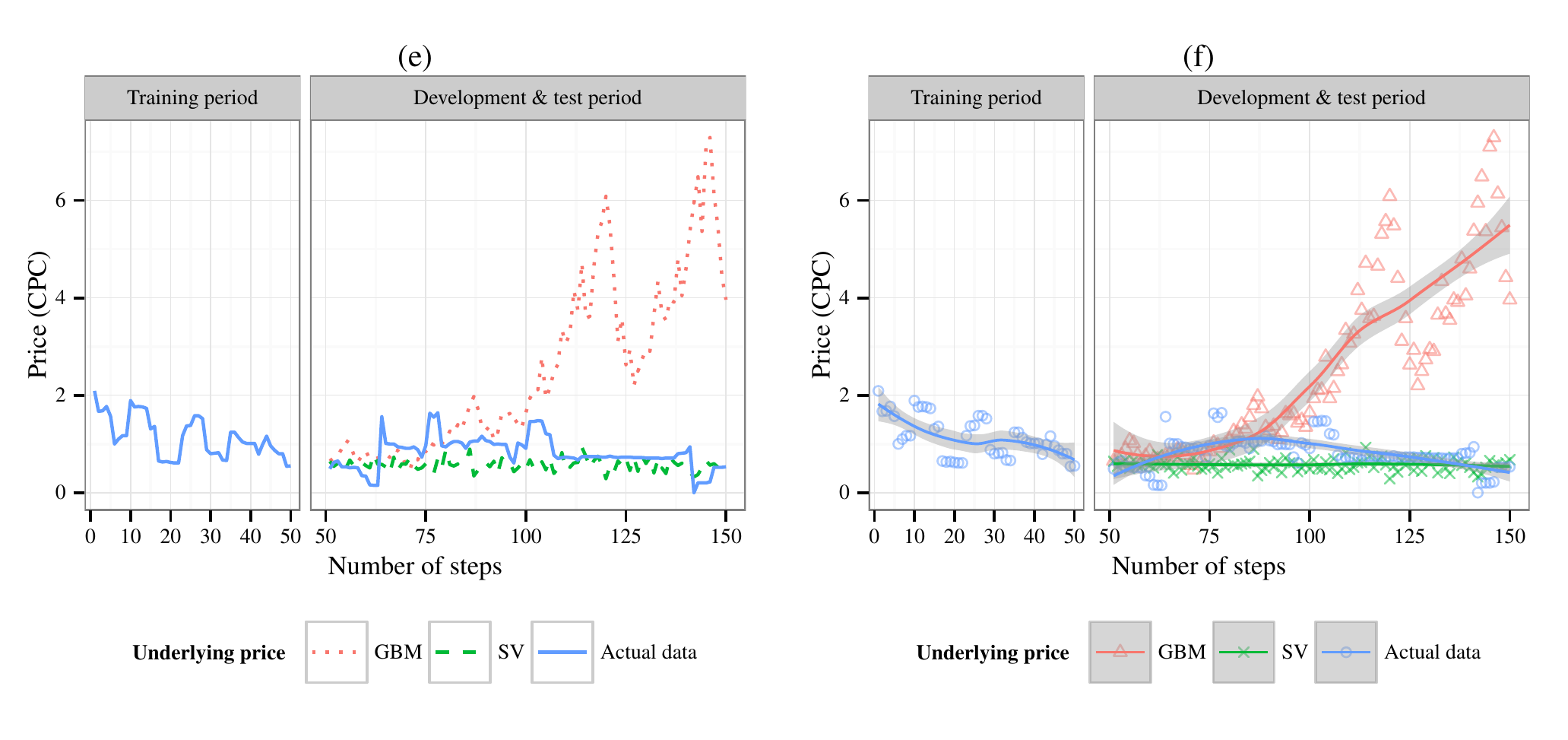}
\caption{Empirical example of comparing the fitness of GBM and SV models for the keyword \lq\lq{}kinect for xbox 360\rq\rq{} from the Google AdWords dataset. The training period is from time step 1 to 50, the development and test periods are from time step 51 to 150. Plot (a), (c), (e) illustrates three instances of simulated paths from the estimated GBM and SV, respectively. Plot (b), (d), (f) provides the corresponding smooth pattern and confidence interval of plot (a), (c), (e).} 
\label{fig:fitness_example} 
\end{figure*}

\begin{table*}[t]
\centering
\caption{Comparing the model fitness for all 31 ad slots in the SSP dataset. L-2 distance is the Euclidean distance, and the number represents the percentage of ad slots which shows that the SV model has a better fitness (i.e., a smaller L-2 distance).} 
\label{tab:fitness_ssp}
\begin{tabular}{r|r|r|r}
\hline
Training set (31 days) & Development \& & L2 distance of & 
L2 distance of \\
 & test set (7 days) & simulated paths & smoothed simulated paths\\
\hline
08/01/2013-07/02/2013 & 08/02/2013-14/02/2013 & 
54.8387\% & 67.7419\% \\
\hline
\end{tabular}
\vspace{5pt}
\caption{Comparing the model fitness for the non-GBM keywords in the Google AdWords dataset. L-2 distance is the Euclidean distance, and the number represents the percentage of non-GBM keywords which shows that the SV model has a better fitness (i.e., a smaller L-2 distance). }
\label{tab:fitness_google}
\begin{tabular}{r|r|r|r|r|r}
\hline
Market &  Group & Training set & Development~\& & L2 distance of & 
L2 distance of \\
 &  & & test set (31 days) & simulated paths & smoothed simulated paths\\
\hline
\multirow{4}{*}{US}  & 1 & 25/01/2012-24/02/2012 & 24/02/2012-25/03/2012 & 82.8571\%  & 80.0000\%    \\
                     & 2 & 30/03/2012-29/04/2012 & 29/04/2012-31/05/2012  & 94.8718\% & 96.1538\% \\
                     & 3 & 10/06/2012-12/07/2012 & 12/07/2012-17/08/2012 & 64.2857\%  & 64.2857\% \\
                     & 4 & 10/11/2012-11/12/2012 & 11/12/2012-10/01/2013 & 98.1481\%  & 100.0000\% \\                    
\hline
\multirow{4}{*}{UK}  & 1 & 25/01/2012-24/02/2012 & 24/02/2012-25/03/2012 & 96.3636\%  &  90.9091\% \\
                     & 2 & 30/03/2012-29/04/2012 & 29/04/2012-31/05/2012 & 98.2456\% & 94.7368\%
\\
                     & 3 & 12/06/2012-13/07/2012 & 13/07/2012-19/08/2012 & 58.0645\%  & 67.7419\%  \\
                     & 4 & 18/10/2012-22/11/2012 & 22/11/2012-24/12/2012 & 72.2222\%  & 80.5556\%  \\                    
\hline
\end{tabular}
\end{table*}
 
Figure~\ref{fig:fitness_example} gives an empirical example showing the model fitness for the situation where the GBM assumption is not valid. Three different instances of simulated paths are generated from the GBM and SV models for the same keyword. Figure~\ref{fig:fitness_example}(a),(c),(d) compares the simulations from these two models with the actual winning payment CPCs in real-time auctions. The smooth movement pattern of these three instances are also examined in Figure~\ref{fig:fitness_example}(b),(d),(f). It is obvious that the SV model has a better fitness to the data. In addition, the Euclidean distance (also called the \emph{L-2 distance}) is used to examine the similarity of the simulated path and the test data. The overall results of the ad slots and keywords in our datasets are presented in Tables~\ref{tab:fitness_ssp}-\ref{tab:fitness_google}, which show that the SV model has a general better fitness to real data. 

\subsection{Validation of the Option Pricing Model}

We now examine the proposed ad option pricing method via two sequential Monte Carlo simulation methods. By using the terminal value pricing formula~\citep{Bjork_2009}, the option price $\pi_0$ can be estimated as follows:
\begin{align}
\pi_0 = & \ \sum_{j = 1}^{\widetilde{n}} \widetilde{r}^{\ -n} 
\bigg( \frac{1}{1000 H} M_j(t_n) - F^C \bigg)^+,
\label{eq:option_price_simulations}
\end{align}
where $M_j(t_n)$ can be generated by either Euler or Milstein discretisation schemes~\citep{Glasserman}:\\
\textbf{Euler Scheme}
\begin{align}
M(t_i + \Delta t) = & \ M(t_i) e^{(r - \frac{1}{2} \sigma^2(t_i)) \Delta t + \sigma(t_i) \sqrt{\Delta t} \epsilon_i}, \label{eq:simulation_euler_1}
\\ 
\sigma(t_i + \Delta t) = & \ \sigma(t_i) + \kappa (\theta - \sigma(t_i)) \Delta t + \delta \sqrt{\sigma(t_i) \Delta t} \varepsilon_i, \label{eq:simulation_euler_2}
\end{align}
\textbf{Milstein Scheme}
\begin{align}
M(t_i + \Delta t) = & \ M(t_i) e^{(r - \frac{1}{2} \sigma^2(t_i)) \Delta t + \sigma(t_i) \sqrt{\Delta t} \epsilon_i},\label{eq:simulation_milstein_1}
\\ 
\sigma(t_i + \Delta t) = & \ \sigma(t_i) + \kappa (\theta - \sigma(t_i)) \Delta t \nonumber \\
& + \delta \sqrt{\sigma(t_i) \Delta t} \varepsilon_i + \frac{1}{4} \delta^2 \Delta t (\varepsilon_i^2 - 1), \label{eq:simulation_milstein_2}
\end{align}
where $\epsilon_i \sim \mathbf{N}(0,1), \varepsilon_i \sim \mathbf{N}(0,1)$.

\begin{figure*}[tp]
\centering
\includegraphics[width=0.825\linewidth]{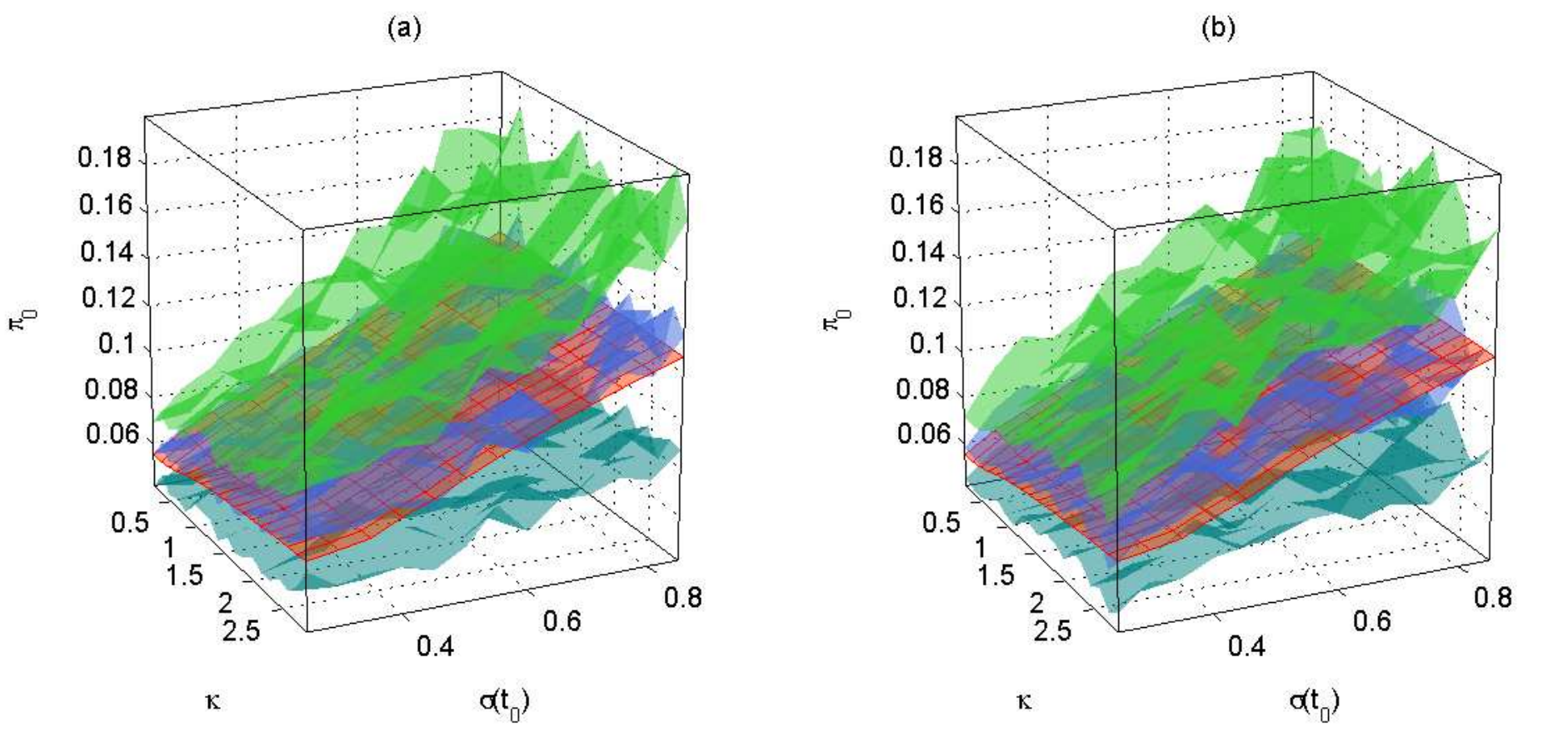}
\includegraphics[width=0.825\linewidth]{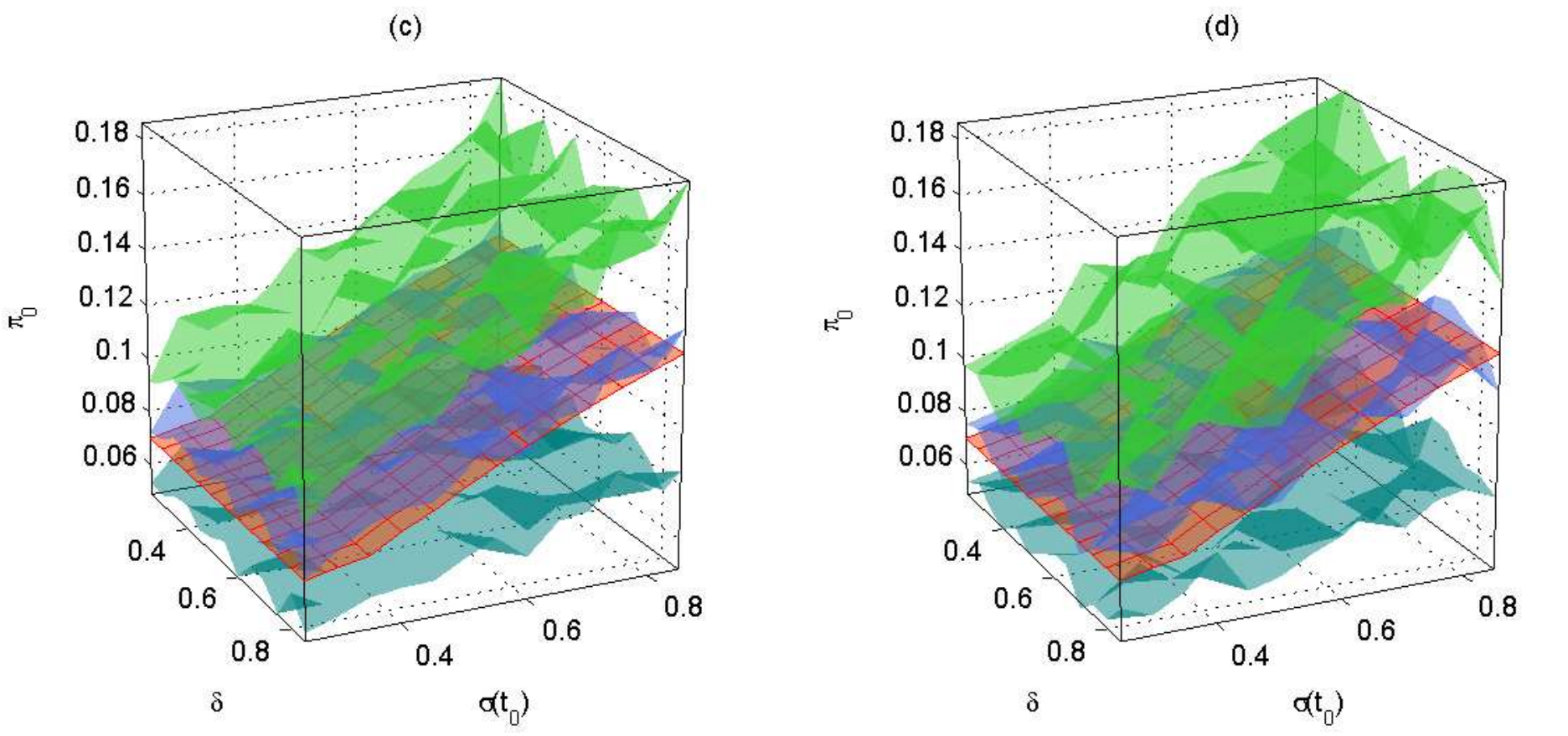}
\includegraphics[width=0.825\linewidth]{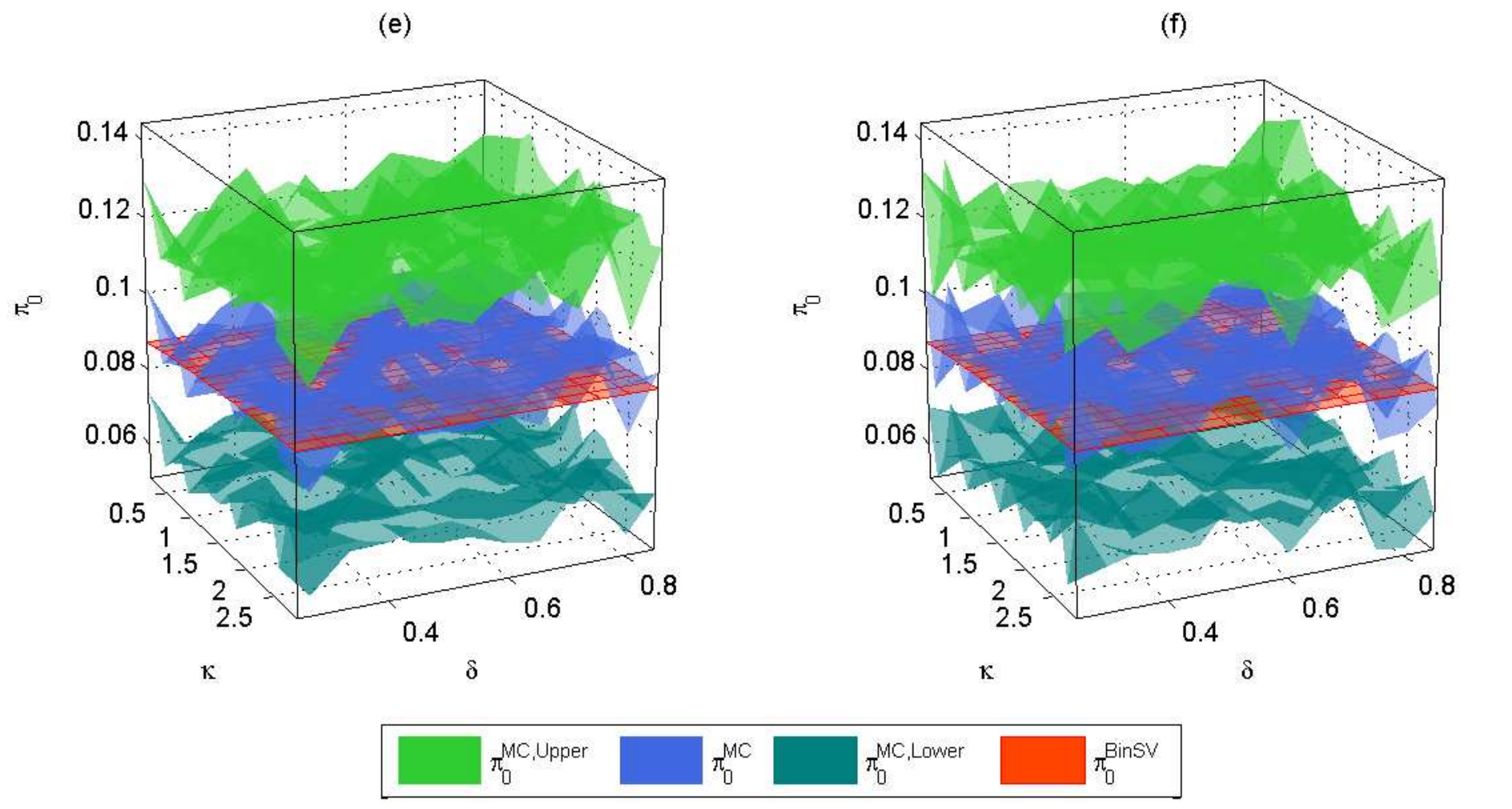}
\caption{Example of model validation tests: (a),(c),(e) Euler scheme; (b),(d),(f) Milstein scheme. The initial values and parameters settings are: $M(t_0)=20, F^C=0.633, r=0.05, \sigma(t_0)=0.5, \kappa = 3, \theta = 0.75, \delta = 0.35$.}
\label{fig:model_validation}
\end{figure*}

These two methods have been widely used in validating the pricing models for exotic options in finance. There are two strong benefits. First, they are developed directly based on the discretisation forms of the underlying dynamics, easy to implement and have good convergence performance to the closed-form solution. Second, they provide a natural criteria for controlling errors. Consider that the errors are controlled with 95\% probability, the following criteria can be used to test the option price calculated from our proposed model:
\begin{align}
\pi_0^{\textbf{BinSV}}
\in \ \ & \Bigg[
\underbrace{\pi_0^{\textbf{MC}} - 1.96 \frac{\widetilde{r}^{\ -n} \textrm{std}(\Phi(M(t_n)))}{\sqrt{\widetilde{n}}}}_{= \pi_0^{\textbf{MC, Lower}}}
, \nonumber \\ 
& \ \ \
\underbrace{\pi_0^{\textbf{MC}} + 1.96 \frac{\widetilde{r}^{\ -n} \textrm{std}(\Phi(M(t_n)))}{\sqrt{\widetilde{n}}}}_{= \pi_0^{\textbf{MC, Upper}}}
\Bigg], \nonumber
\end{align}
where $\pi_0^{\textbf{BinSV}}$ represents the option price calculated from our proposed censored binomial lattice method, $\pi_0^{\textbf{MC}}$ represents the option price calculated from Monte Carlo simulations, $\pi_0^{\textbf{MC, Lower}}$ and $\pi_0^{\textbf{MC, Upper}}$ represent the lower and upper bounds of $\pi_0^{\textbf{MC}}$.

Figure~\ref{fig:model_validation} provides our model validation test. We price an ad option using the proposed censored binomial lattice and the discussed two Monte Carlo simulation methods respectively. The model parameters are changed in certain intervals against each other in order to investigate the sensitivity of the calculated option price to the values of parameters. It is not difficult to see that our proposed lattice method is robust and accurate because $\pi_0^{\textbf{BinSV}}$ is very close to $\pi_0^{\textbf{MC}}$ and always lies in the confidence interval for different model parameters' values.

\begin{sidewaystable*}
\centering
\caption{Empirical example of an advertiser's delivery of an ad slot from the SSP dataset in RTB (Note: CTR is $0.03$ and the non-integer numbers are displayed at 4 digits after the decimal point while in computing we consider 25-digit scale).}
\label{tab:example_delivery_1}
\begin{tabular}{r|r|rrrrrrr|r}
\hline
Day  & 0 & 1 & 2 & 3 & 4 & 5 & 6 & 7 & Total \\
\hline
Date & 07/02/2013 & 08/02/2013 & 09/02/2013 & 10/02/2013 & 11/02/2013 & 12/02/2013 
& 13/02/2013 & 14/02/2013 & \\
Average payment CPM  
&  0.7427 & 0.9585 & 0.9770 & 0.9666 & 0.8754 & 0.8513 & 0.8294 & 0.9903 &   \\
No. of total impressions generated & & 8298  & 8277  & 8190   &  7971  & 8097  &  8201 &  3812 & 52846 \\
\hline
Budget    &    & 5.0000  & 5.0000  & 5.0000  & 5.0000  & 5.0000  & 5.0000  & 5.0000  & 35.0000\\
No. of impressions received   &     & 5210   &  5113  & 5166  & 5711  & 5867  & 6028 & 3812 & 36907\\
No. of clicks received   &    &  156  & 153 &  154  & 171 &  176 &  180 &  114 & 1104 \\
Used budget &      &  5.0000  &  5.0000 & 5.0000 & 4.9830  & 5.0000  &  4.9926  &  3.7748 &
33.7504\\
\hline
\end{tabular}
\vspace{40pt}
\caption{Empirical example of an advertiser's delivery of buying ad options for an advertisement slot in the SSP dataset (Note: CTR is $0.03$ and the non-integer numbers are displayed at 4 digits after the decimal point while in computing we consider 25-digit scale).}	
\label{tab:example_delivery_2}
\begin{tabular}{r|rrr|rrrrrrrrrr}
\hline
Day & Date & Average & No. of total & Budget & Remaining & No. of & Expiration & Option & Strike & No. of & No. of & No. of & Used \\
 & & payment &  impressions & & budget & options & date & price & price & options & impressions & clicks & budget \\
 & & CPM & generated & & &   & & & CPC &  exercised & received & received &  \\
\hline
0 & 07/02/2013 & 0.7427 & & &  & 201 & 08/02/2013 & 0.0025 & 0.0223 &  & & & 0.4982 \\
  &            & 0.7427 & & &  & 201 & 09/02/2013 & 0.0025 & 0.0223 &  & & & 0.4988 \\
  &            & 0.7427 & & &  & 201 & 10/02/2013 & 0.0025 & 0.0223 &  & & & 0.5117 \\
  &            & 0.7427 & & &  & 201 & 11/02/2013 & 0.0026 & 0.0223 &  & & & 0.5192 \\
  &            & 0.7427 & & &  & 200 & 12/02/2013 & 0.0026 & 0.0223 &  & & & 0.5271 \\
  &            & 0.7427 & & &  & 200 & 13/02/2013 & 0.0027 & 0.0223 &  & & & 0.5379 \\
  &            & 0.7427 & & &  & 199 & 14/02/2013 & 0.0027 & 0.0223 &  & & & 0.5427 \\
\hline  
1 & 08/02/2013 & 0.9585 & 8298 & 5.0000  & 4.5018 & 201 & 08/02/2013 & 0.0022 & 0.0223 & 201 & 6816 & 204 & 4.5013 \\
2 & 09/02/2013 & 0.9770 & 8277 & 5.0000 & 4.5012 &  201 & 09/02/2013 & 0.0021 & 0.0223 & 201 & 6770 & 203 & 4.5011 \\
3 & 10/02/2013 & 0.9666 & 8190 & 5.0000  & 4.4883 & 201 & 10/02/2013 & 0.0019 & 0.0223 & 201 & 6742 & 202 & 4.4878 \\
4 & 11/02/2013 & 0.8754 & 7971 & 5.0000 & 4.4808 & 201 & 11/02/2013 & 0.0019 & 0.0223 & 199 & 6836 & 205 & 4.4298 \\
5 & 12/02/2013 & 0.8513 & 8097 & 5.0000 & 4.4729 & 200 & 12/02/2013 & 0.0018 & 0.0223 & 200 & 6776 & 203 & 4.4727 \\
6 & 13/02/2013 & 0.8294 & 8201 & 5.0000 & 4.4621 & 200 & 13/02/2013 & 0.0017 & 0.0223 & 197 & 6792 & 204 & 4.4616  \\
7 & 14/02/2013 & 0.9903 & 3812 & 5.0000 & 4.4573 & 199 & 14/02/2013 & 0.0017 & 0.0223 & 114 & 3812 & 114 & 2.5463  \\
\hline
Total &  & & 52846 & 35.0000 & & &&&  & & 44544 & 1335 & 33.0362\\
\hline
\end{tabular}
\end{sidewaystable*}

\subsection{Delivery Performance for Advertiser}

Tables~\ref{tab:example_delivery_1}-\ref{tab:example_delivery_2} present an empirical example that compares an advertiser's delivery performance between RTB and ad options. Table~\ref{tab:example_delivery_1} shows the advertiser's delivery performance in RTB with a fixed daily budget. If the supplied impressions are at same levels and if the average winning payment CPMs increase, the advertiser will receive fewer impressions. In Table~\ref{tab:example_delivery_2}, the advertiser buys several ad options in advance. Consider if he purchases an ad option with expiration date 08/02/2013, he has the right to secure impressions that will be created on 08/02/2013 at a fixed CPC. Here the advertiser is assumed to use his daily budget from the corresponding delivery date to pay the upfront option price. Hence, as shown in Table~\ref{tab:example_delivery_2}, the advertiser's strategy is to purchase as many options as possible, and the remaining daily budgets will be used on the corresponding delivery dates. We use the actual bids from RTB to simulate the real-time feeds of the spot market, so if the market value of a click is higher than the fixed payment, the advertiser will use ad options to secure the needed clicks and then pay the fixed CPCs accordingly. Otherwise, the advertiser will obtain the equivalent clicks from RTB. Our example shows a \lq\lq{}bull market\rq\rq{} where the average spot CPM in the test set is far higher than the initial CPM. Therefore, ad options would be actively used by the advertiser to purchase the clicks. Compared to Table~\ref{tab:example_delivery_1}, the advertiser can receive more clicks (increased by 20.92\%) in a bull market via ad options.

\begin{table*}[t]
\centering
\caption{Overview of the improvement in delivery performance by using ad options for all ad slots in the SSP dataset.}
\label{tab:overview_delivery_ssp}
\begin{tabular}{r|rr}
\hline
			 			  			        & Bull market & Bear market \\
\hline 
Change on used budget (\%) 			        & -8.7878\% & --	\\
Change on delivery of impressions (\%)  	& 6.1781\%  & --    \\
\hline
\end{tabular}
\vspace{10pt}
\caption{Overview of the improvement in delivery performance by using ad options for keywords in the Google AdWords dataset.}
\label{tab:overview_delivery_google}
\begin{tabular}{r|r|r|r|r|r}
\hline
\multirow{2}{*}{Market}	& \multirow{2}{*}{Group} & \multicolumn{2}{l|}{Change in used budget (\%)} & \multicolumn{2}{l}{Change in delivery of impressions (\%)} \\
\cline{3-6}
& & Bull market & Bear market & Bull market & Bear market \\
\hline 
\multirow{4}{*}{US} & 1 & 0.3447\% & 2.3438\% & 9.3050\%  & -0.1122\%	\\
   & 2 & 1.7748\% & 3.9687\% & 2.3153\%  & -2.6285\%  	\\
   & 3 & 0.5372\% & 4.8567\% & 44.3735\% & -0.0940\% 	\\
   & 4 & 5.6288\% & 29.3626\%   & 1.6433\% &  -1.0993\% \\
\hline
\multirow{4}{*}{UK} & 1 & 21.4285\% & 6.8940\% & 3.0717\% & -0.2523\%  \\
   & 2 & 5.4426\%  &  0.0000\%  &  0.4419\%  &   0.0000\%  \\
   & 3 & 10.9285\% &  3.8474\% &  28.7706\% &  -2.1066\% \\
   & 4 & 6.7155\%  & 0.1552\%  & 16.6955\% &  -2.1550\% \\
\hline
\end{tabular}
\end{table*}

Similar experiments are conducted for all ad slots in our datasets. The overall results are presented in Tables~\ref{tab:overview_delivery_ssp}-\ref{tab:overview_delivery_google}. For the SSP dataset, we consider the ad options that allow advertisers to pay a fixed CPC to purchase impressions of targeted ad slots. For the Google dataset, we consider the ad options that allow advertisers to pay a fixed CPM to purchase clicks of their targeted keywords. To summarize, we find that an advertiser's daily budget can be used more effectively in a bull market and that his delivery increases as well. The advertiser's average cost spent on each impression or click is reduced. In a bear market (i.e., the underlying price decreases), the advertiser will use the ad options less (and sometimes not at all) and the maximum cost is just the option price. It is worth noting that here we consider the ad options are in the money at time $0$ (i.e., the strike price is less than the current underlying price). In Table~\ref{tab:example_delivery_1}, there are 4 ad slots that exhibit somewhat bear markets. However, these 4 ad slots do not receive enough bids in the test set and the actual winning payment CPMs are just around its floor reserve level (i.e., the CPM is \pounds 0.01 so the per impression price is \pounds 0.00001). Since these prices will seriously bias the results, we do not take them into account in the situation of a bear market.

\subsection{Revenue Analysis for Publisher and Search Engine}

\begin{figure*}[t]
\centering
\includegraphics[width=0.825\linewidth]{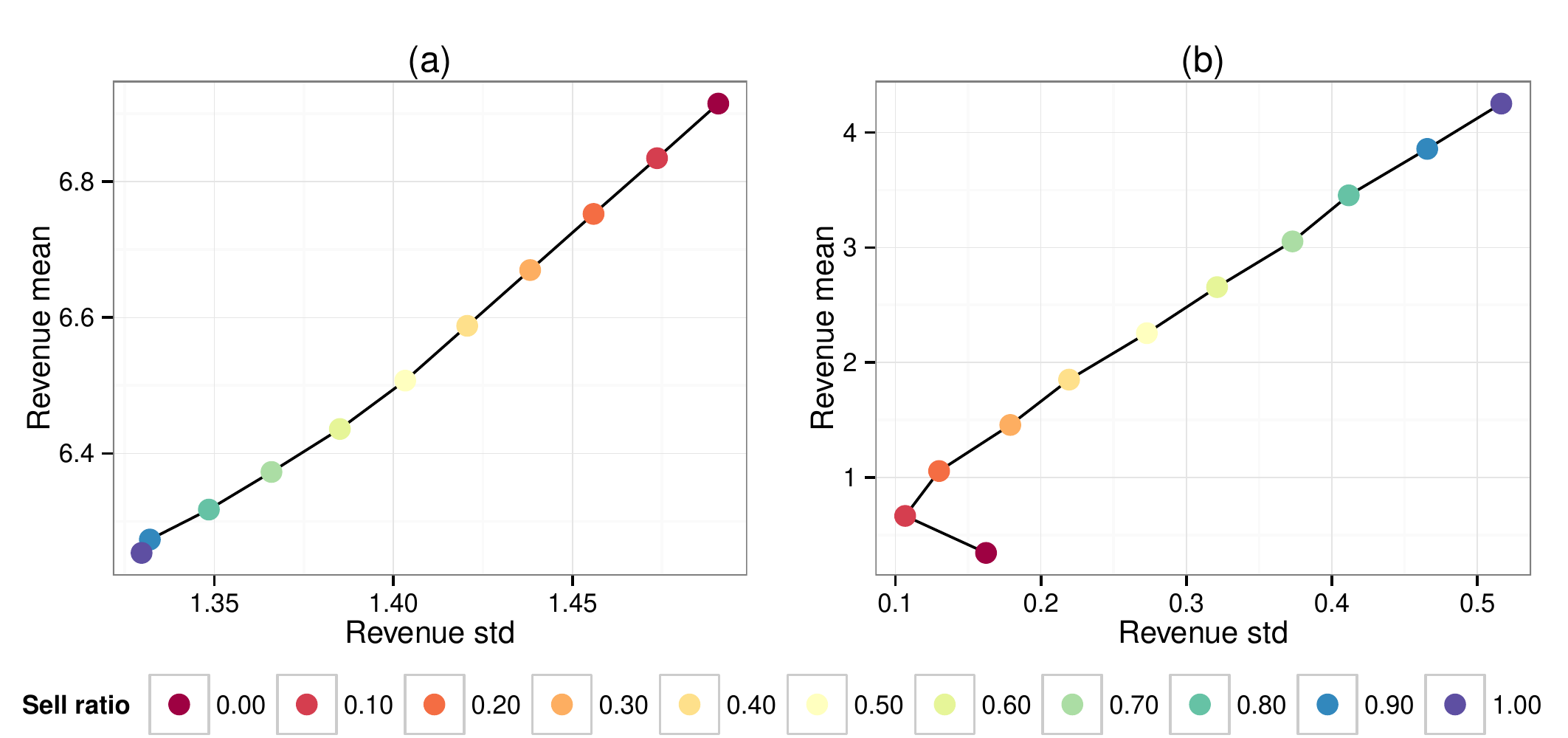}
\caption{Empirical examples of the publisher's revenue: (a) from an ad slot in the bull market; and (b) from an ad slot in the bear market. The sell ratio represents the percentage of future daily impressions that are sold in advance via display ad options. Note that here the ad slot in the bear market does not receive enough bids in the test set, so we randomly simulate some underlying prices for the bear market.}
\label{fig:example_revenue_ssp}
\end{figure*}

We also investigate the revenue effects when a certain amount of future impressions or clicks can be sold in advance. Figure~\ref{fig:example_revenue_ssp} provides two empirical examples of ad slots from the SSP dataset: one exhibits the bull market while the other shows the bear market. The sell ratio in the figure represents the percentage of future impressions that are sold in advance via ad options; therefore, when the sell ratio equals zero, the publisher auctions off all of the future impressions in RTB. Figure~\ref{fig:example_revenue_ssp}(a) suggests that the publisher should sell less future impressions in advance if the future market is bull. This is because ad options will be exercised by advertisers in the future and the obtained revenues from the fixed payment are less than these impressions' market values. Of course, the publisher can choose a certain percentage of future impressions to sell according to his level of risk tolerance or to meet other business objectives. For example, the publisher may be willing to sacrifice some revenues in order to increase the advertisers' engagement in the long run. Conversely, in a bear market, as shown in Figure~\ref{fig:example_revenue_ssp}(b), the publisher is advised to sell more future impressions in advance because there is more upfront income if more display ad options are sold, and in the future advertisers will not exercise the sold options. Therefore, the increased revenue comes from the option price. 

\begin{table*}[tp]
\centering
\caption{Overview of the improvement in revenue by selling display ad options for ad slots in the SSP dataset.}
\label{tab:overview_revenue_ssp} 
\begin{tabular}{r|rr}
\hline
			 			  			        & Bull market & Bear market \\
\hline 
Change on mean (\%) 			        &  -7.1283\%  &  726.3085\%  \\
Change on standard deviation (\%)  	    &  -2.7041\%  &  196.0547\%   \\
\hline
\end{tabular}
\vspace{10pt}
\caption{Overview of the improvement in revenue by selling display ad options for ad slots in the Google AdWords dataset. }
\label{tab:overview_revenue_google}
\begin{tabular}{r|r|r|r|r|r}
\hline
\multirow{2}{*}{Market}	& \multirow{2}{*}{Group} & \multicolumn{2}{l|}{Change in mean (\%) } & \multicolumn{2}{l}{Change in standard deviation (\%) } \\
\cline{3-6}
&  &  Bull market & Bear market & Bull market & Bear market \\
\hline 
\multirow{4}{*}{US} 
   & 1 & -20.5880\% &  22.3898\% &  -0.6507\% &   9.3291\%  \\
   & 2 & -23.2971\% &  17.1898\% &  -17.6508\% &  9.4175\%  \\
   & 3 & -32.8388\% &   69.9113\% &  -21.9468\% &   -2.1065\% \\
   & 4 & -24.4710\% &   8.9650\% &  -10.6024\% &  95.4868\%  \\ 
\hline
\multirow{4}{*}{UK}
   & 1 & -8.5463\% 	 &  15.4155\% &  4.5617\% &   10.4116\%    \\
   & 2 & -20.0632\%  &   4.3816\% &  -16.0239\%  &  6.8847\%  \\
   & 3 & -16.9050\%  &  30.7737\% &  -11.4811\% &  -19.4625\%  \\
   & 4 & -21.8142\%  &   7.6342\% &  -19.4368\% &  0.3877\%  \\
\hline
\end{tabular}
\end{table*}

Based on the above analysis, we examine the revenue effects across all ad slots and keywords in our datasets. In the experiments, the display ad options in a bull market are priced in the money while in a bear market they are priced out of the money. The sell ratio is set at 0.20 in a bull market while it is set at 0.80 in a bear market. The overall results are presented in Tables~\ref{tab:overview_revenue_ssp}-\ref{tab:overview_revenue_google}, which further confirm our analysis in the empirical examples. The average revenue is reduced in the bull market as well as the standard deviation (i.e., one kind of revenue risk). However, as described, the publisher (or search engine) may be willing to sacrifice some revenue to establish a long-term relationship with advertisers. In a bear market, the average revenue increases significantly. This is because fewer display ad options are exercised. Many premium advertisers join RTB so that the market equilibrium is almost as same as that in an environment with only auctions. Finally, the publisher (or search engine) earns the upfront payment without providing guaranteed deliveries.

\section{Concluding Remarks}
\label{ao:conclusion}

This paper described a new ad option tailored to the unique environment of display advertising. A binomial lattice framework with censored probabilities was proposed to price the ad option where the underlying prices follow a SV model. We also reviewed and examined several lattice methods for pricing the ad option with the GBM underlying model. Our developments were examined and validated by experiments using real advertising data. 

We believe that the proposed ad options will soon be welcomed by display advertising market. Several similar but different developments appeared are able to support our point of view. They are: \vspace{-5pt}
\begin{description}
\item[\rm 09/2013] AOL's Programmatic Upfront\footnote{\href{http://www.aolplatforms.com}{www.aolplatforms.com}}.\vspace{-5pt}
\item[\rm 03/2013] OpenX Programmatic Guarantee\footnote{\href{http://openx.com/whitepaper/programmatic-premium-current-practices-and-future-trends}{www.openx.com}
}.\vspace{-5pt}
\item[\rm 10/2012] Adslot Media's Programmatic Direct Media Buying\footnote{\href{http://www.automatedguaranteed.com}{www.automatedguaranteed.com}}.\vspace{-5pt}
\item[\rm 10/2012] Shiny Ads Direct's End-to-end Programmatic Direct Advertising Platform\footnote{\href{www.shinyads.com}{www.shinyads.com}}.\vspace{-5pt}
\item[\rm 10/2012] iSOCKET's Programmatic Direct\footnote{\href{www.isocket.com}{www.isocket.com}}.\vspace{-5pt}
\end{description}

Our work differs to the above developments in many aspects. First, the proposed ad options provide flexible guaranteed deliveries (e.g., no obligation of exercise, choosing the fixed payment that is different to the underlying inventory measurement model) while other recent developments do not provide such features. Second, we proposed a generalized pricing model which can deal with those situations when the GBM model fails.

There are three major limitations of the study in this paper, which can be further explored for future research. Firstly, we did not explicitly consider the capacity issue in option pricing. Therefore, there may exist the situations that a publisher or search engine can not guarantee the delivery of impressions or clicks sold by options. In our current study, we consider the seller has a good estimation of the inventories that will be created in the future and rationally sells the future inventories in advance via options. If the seller over sells the future inventories, we also assume that he can buy some similar inventories on the spot market once the option buyers request to exercise the options. In such case, the revenues of the seller will decrease. The capacity issue is an interesting topic to further discuss in details because it has two challenges. The first challenge is to price an ad option with explicitly considering the estimation of future supply and demand of inventories, where the latter two variables can be described to be static~\citep{Wang_2012_1} or dynamic like the Poisson process~\citep{Gallego_1994}. The second challenge is considering the penalty into option pricing. If the seller fails to deliver inventories requested by option holders, the seller should pay a certain amount of penalty fee~\citep{Chen_2014_2}. However, with the penalty setting, some advertisers who only pursue the penalty may game the system~\citep{Constantin_2009}, which will further affect the calculated option price, and such effect will also generate some scenarios like the implied volatility in financial market. The second limitation is that the proposed model can not capture the jumps and volatility clusters of underlying inventory prices. It might be of interest to discuss these stylized facts in ad option pricing. The third limitation is the zero correlation of the two standard Brownian motions in our proposed dynamics. If their correlation is not zero, the option pricing would be more sophisticated under the lattice framework. \cite{Heston_1993} proposed a good solution in the continuous-time settings, which can also be extended to solve our problem in online advertising.

\bibliography{mybib}

\begin{thebibliography}{40}
\expandafter\ifx\csname natexlab\endcsname\relax\def\natexlab#1{#1}\fi
\expandafter\ifx\csname url\endcsname\relax
  \def\url#1{\texttt{#1}}\fi
\expandafter\ifx\csname urlprefix\endcsname\relax\def\urlprefix{URL }\fi

\bibitem[{Bachelier(1900)}]{Bachelier_1900}
Bachelier, L., 1900. Th\'{e}orie de la sp\'{e}culation. Annales Scientifiques
  de l'\'{E}cole Normale Sup\'{e}rieure 3~(17), 21--86.

\bibitem[{Bharadwaj et~al.(2010)Bharadwaj, Ma, Schwarz, Shanmugasundaram, Vee,
  Xie, and Yang}]{Bharadwaj_2010}
Bharadwaj, V., Ma, W., Schwarz, M., Shanmugasundaram, J., Vee, E., Xie, J.,
  Yang, J., 2010. Pricing guaranteed contracts in online display advertising.
  In: Proceedings of the 19th ACM International Conference on Information and
  Knowledge Management. ACM, Toronto, ON, Canada, pp. 399--408.

\bibitem[{Bj{\"{o}}rk(2009)}]{Bjork_2009}
Bj{\"{o}}rk, T., 2009. Arbitrage Theory in Continuous Time, 3rd Edition. Oxford
  University Press.

\bibitem[{Black and Scholes(1973)}]{Black_1973}
Black, F., Scholes, M., 1973. The pricing of options and corporate liabilities.
  Journal of Political Economy 81~(3), 637--654.

\bibitem[{Boer(2002)}]{Boer_2002}
Boer, P., 2002. The Real Options Solution Finding Total Value in a High-Risk
  World. John Wiley.

\bibitem[{Boyle(1986)}]{Boyle_1986}
Boyle, P., 1986. Option valuation using a three-jump process. International
  Options Journal 3, 7--12.

\bibitem[{Boyle(1988)}]{Boyle_1988}
Boyle, P., 1988. A lattice framework for option pricing with two state
  variables. Journal of Financial and Quantitative Analysis 23~(1), 1--12.

\bibitem[{Chen et~al.(2015)Chen, Wang, Cox, and Kankanhalli}]{Chen_2015_1}
Chen, B., Wang, J., Cox, I., Kankanhalli, M., October 2015. Multi-keyword
  multi-click advertisement option contracts for sponsored search. ACM
  Transactions on Intelligent Systems and Technology 7~(1).

\bibitem[{Chen et~al.(2014)Chen, Yuan, and Wang}]{Chen_2014_2}
Chen, B., Yuan, S., Wang, J., 2014. A dynamic pricing model for unifying
  programmatic guarantee and real-time bidding in display advertising. In:
  Proceedings of the 8th International Workshop on Data Mining for Online
  Advertising. ACM, New York, NY, USA, pp. 1--9.

\bibitem[{Constantin et~al.(2009)Constantin, Feldman, Muthukrishnan, and
  P\'{a}l}]{Constantin_2009}
Constantin, F., Feldman, J., Muthukrishnan, S., P\'{a}l, M., 2009. An online
  mechanism for ad slot reservations with cancellations. In: Proceedings of the
  20th Annual ACM-SIAM Symposium on Discrete Algorithms. SIAM, New York
  Marriott Downtown, NY, USA, pp. 1265--1274.

\bibitem[{Cox et~al.(1979)Cox, Ross, and Rubinstein}]{Cox_1979}
Cox, J., Ross, S., Rubinstein, M., 1979. Option pricing: {a} simplified
  approach. Journal of Financial Economics 7, 229--263.

\bibitem[{Edelman et~al.(2007)Edelman, Ostrovsky, and Schwarz}]{Edelman_2007_2}
Edelman, B., Ostrovsky, M., Schwarz, M., 2007. {Internet} advertising and the
  generalized second-price auction: {selling} billions of dollars worth of
  keywords. American Economic Review 97~(1), 242--259.

\bibitem[{Florescu and Viens(2005)}]{Florescu_2005}
Florescu, I., Viens, F., 2005. A binomial tree approach to stochastic
  volatility driven model of the stock price. Annals of University of Craiova
  32, 126--142.

\bibitem[{Florescu and Viens(2008)}]{Florescu_2008}
Florescu, I., Viens, F., 2008. Stochastic volatility: {o}ption pricing using a
  multinomial recombining tree. Applied Mathematical Finance 15~(2), 151--181.

\bibitem[{Gallego and {van Ryzin}(1994)}]{Gallego_1994}
Gallego, G., {van Ryzin}, G., 1994. Optimal dynamic pricing of inventories with
  stochastic demand over finite horizons. Management Science 40~(8), 999--1020.

\bibitem[{Glasserman(2003)}]{Glasserman}
Glasserman, P., 2003. Monte Carlo Methods in Financial Engineering. Springer.

\bibitem[{{Google}(2011)}]{Google_2011}
{Google}, 2011. The arrival of real-time bidding and what it means for media
  buyers. White Paper.

\bibitem[{Haahtela(2010)}]{Haahtela_2010}
Haahtela, T., 2010. Recombining trinomial tree for real option valuation with
  changing volatility. DOI: \url{http://dx.doi.org/10.2139/ssrn.1932411}.

\bibitem[{Heston(1993)}]{Heston_1993}
Heston, S., 1993. A closed-form solution for. options with stochastic
  volatility with applications to bond and currency options. The Review of
  Financial Studies 6~(2), 327--343.

\bibitem[{{Interactive Advertising Bureau of Canada}(2015)}]{IAB_2015}
{Interactive Advertising Bureau of Canada}, 2015. Real-time bidding glossary.
  \url{http://iabcanada.com/files/RTB-Glossary-English.pdf}.

\bibitem[{Kamrad and Ritchken(1991)}]{Kamrad_1991}
Kamrad, B., Ritchken, P., 1991. Multinomial approximating models for options
  with k state variables. Management Science 37~(23), 1640--1652.

\bibitem[{Ljung and Box(1978)}]{Ljung_1978}
Ljung, G., Box, G., 1978. On a measure of lack of fit in time series models.
  Biometrika 65~(2), 297--303.

\bibitem[{Marathe and Ryan(2005)}]{Marathe_2005}
Marathe, R., Ryan, S., 2005. On the validity of the geometric {Brownian} motion
  assumption. The Engineering Economist 50, 159--192.

\bibitem[{Marshall(2012)}]{Marshall_2012}
Marshall, J., 2012. Valuation of multiple exercise options. Ph.D. thesis,
  University of Western Ontario.

\bibitem[{Merton(1973)}]{Merton_1973}
Merton, R., 1973. The theory of rational option pricing. Bell Journal of
  Economics and Management Science 4~(1), 141--183.

\bibitem[{Moon and Kwon(2010)}]{Moon_2010}
Moon, Y., Kwon, C., 2010. Online advertisement service pricing and an option
  contract. Electronic Commerce Research and Applications 10~(1), 38--48.

\bibitem[{Nelson and Ramaswamy(1990)}]{Nelson_1990}
Nelson, D., Ramaswamy, K., 1990. Simple binomial processes as diffusion
  aapproximation in financial models. The Review of Finanical Studies.

\bibitem[{Primbsa et~al.(2007)Primbsa, Rathinamb, and Yamadac}]{Primbsa_2007}
Primbsa, J., Rathinamb, M., Yamadac, Y., 2007. Option pricing with a
  pentanomial lattice model that incorporates skewness and kurtosis. Applied
  Mathematical Finance, 1--17.

\bibitem[{Samuelson(1965)}]{Samuelson_1965_2}
Samuelson, P., 1965. Rational theory of warrant pricing. Industrial Management
  Review 6, 13--31.

\bibitem[{Shapiro and Wilk(1965)}]{Shapiro_1965}
Shapiro, S., Wilk, M., 1965. An analysis of variance test for normality.
  Biometrika 52~(3), 591--611.

\bibitem[{Sharpe(1978)}]{Sharpe_1978}
Sharpe, W., 1978. Investments, 1st Edition. Englewood Cliffs.

\bibitem[{Tian(1993)}]{Tian_1993}
Tian, Y., 1993. A modified lattice approach to option pricing. The Journal of
  Futures Markets.

\bibitem[{Tsay(2005)}]{Tsay_2005}
Tsay, R., 2005. Analysis of Financial Series, 2nd Edition. John Wiley.

\bibitem[{Varian(1987)}]{Varian_1987}
Varian, H., 1987. The arbitrage principle in financial economics. Journal of
  Economic Perspectives 1~(2), 55--72.

\bibitem[{Wang and Chen(2012)}]{Wang_2012_1}
Wang, J., Chen, B., 2012. Selling futures online advertising slots via option
  contracts. In: Proceedings of the 21st International Conference on World Wide
  Web. ACM, Lyon, France, pp. 627--628.

\bibitem[{Wilmott(2006)}]{Wilmott_2006_1}
Wilmott, P., 2006. Paul Wilmott On Quantitative Finance, 2nd Edition. John
  Wiley.

\bibitem[{Yuan and Wang(2012)}]{Yuan_2012}
Yuan, S., Wang, J., 2012. Sequential selection of correlated ads by {POMDPs}.
  In: Proceedings of the 21st ACM International Conference on Information and
  Knowledge Management. ACM, Maui, HI, USA, pp. 515--524.

\bibitem[{Yuan et~al.(2014)Yuan, Wang, Chen, Mason, and Seljan}]{Yuan_2014}
Yuan, S., Wang, J., Chen, B., Mason, P., Seljan, S., 2014. An empirical study
  of reserve price optimisation in real-time bidding. In: Proceedings of the
  20th ACM SIGKDD Conference on Knowledge Discovery and Data Mining. ACM, New
  York, NY, USA, pp. 1897--1906.

\bibitem[{Yuan et~al.(2013)Yuan, Wang, and Zhao}]{Yuan_2013_2}
Yuan, S., Wang, J., Zhao, X., 2013. Real-time bidding for online advertising:
  {m}easurement and analysis. In: Proceedings of the 7th Workshop on Data
  Mining for Online Advertising. ACM, Chicago, IL, USA.

\bibitem[{Zhang(1998)}]{Zhang_1998}
Zhang, P., 1998. Exotic Options, 2nd Edition. World Scientific.

\end{thebibliography}
\bibliographystyle{elsarticle-harv}

\end{document}